\documentclass[a4paper,11pt]{article}
\usepackage{jheppub} 
\usepackage{lineno}
\usepackage{subcaption}
\usepackage{graphicx}
\usepackage{tcolorbox}
\usepackage{dcolumn}

\newcommand{\be}{\begin{equation}}
\newcommand{\ee}{\end{equation}}

\newcommand{\bg}{\begin{gathered}}
\newcommand{\eg}{\end{gathered}}

\usepackage{xcolor}

\title{\boldmath Dissipative fracton superfluids
}

\author[1,2]{Aleksander G\l{}\'{o}dkowski,}\author[1]{Francisco Pe\~na-Ben\'itez,}\author[1]{Piotr Sur\'{o}wka}

\affiliation[1]{Institute of Theoretical Physics, Wroc\l{}aw  University  of  Science  and  Technology,  50-370  Wroc\l{}aw,  Poland}
\affiliation[2]{Max Planck Institute for the Physics of Complex Systems, 01187 Dresden, Germany} 
\emailAdd{aleksander.glodkowski@pwr.edu.pl}
\emailAdd{francisco.pena-benitez@pwr.edu.pl}
\emailAdd{piotr.surowka@pwr.edu.pl}

\abstract{
We present a comprehensive study of hydrodynamic theories for superfluids with dipole symmetry. Taking diffusion as an example, we systematically construct a hydrodynamic framework that incorporates an intrinsic dipole degree of freedom in analogy to spin density in micropolar (spinful) fluids. Subsequently, we study a dipole condensed phase and propose a model that captures the spontaneous breaking of the $U(1)$ charge. The theory explains the role of the inverse Higgs constraint for this class of theories, and naturally generates the gapless field.
Next, we introduce finite temperature theory using the Hamiltonian formalism and study the hydrodynamics of ideal fracton superfluids. Finally, we postulate a derivative counting scheme and incorporate dissipative effects using the method of irreversible thermodynamics. We verify the consistency of the dispersion relations and argue that our counting is systematic.  
}
\usepackage{etoolbox}
    \makeatletter
    \patchcmd{\maketitle}{\@fpheader}{}{}{}
    \makeatother

\begin{document}
\maketitle

\section{Introduction}

Finite temperature and finite density phenomena, in which the microscopic constituents interact strongly, are well-captured by hydrodynamics. In consequence, the hydrodynamic paradigm is a universal and systematic way of organizing the effective low energy degrees of freedom in strongly coupled matter, e.g. see the quark-gluon plasma \cite{Schenke:2021mxx}, and electrons in graphene \cite{narozhny_hydrodynamic_2022}.
Thus, the study of strongly coupled field theories through the lens of hydrodynamics presents a rich and fertile ground for advancing our understanding of complex interacting systems. Following this path, we are able to access non-perturbative effects, which allows us to analyze their physical consequences, that are otherwise inaccessible due to the computational and conceptual challenges posed by strong coupling. Moreover, this approach has the potential to unveil new aspects of quantum field theory and may lead to novel applications in materials science and quantum technologies.

This work concentrates on examining the long-wavelength characteristics of many-body systems that display dipole symmetry. Employing hydrodynamics we explore the domain of fracton superfluids, i.e. many-body systems breaking spontaneously dipole conserving symmetry group. Our investigation is driven by a dual purpose. Firstly, it aims to enrich our understanding of hydrodynamics itself, especially in its application to systems with unconventional symmetries. For example, the linear realization of the dipole symmetry group on a scalar field requires the theory to be non-Gaussian and  strongly coupled \cite{gauge,Jensen:2022iww,Molina-Vilaplana:2023doq}. Secondly, we endeavor to identify unique transport characteristics inherent to dipole-conserving systems. This aspect holds promise for future experimental validation, particularly in systems manifesting emergent dipole symmetry, such as tilted Fermi-Hubbard chains \cite{Guardado_Sanchez_2020,Gromov_hydro_2020,Nandy:2023bop} or assemblies of topological defects \cite{pretko_fracton-elasticity_2018,pretko_crystal--fracton_2019,radzihovsky_fractons_2020,radzihovsky_quantum_2020,gromov_duality_2020,surowka_dual_2021,hirono_effective_2022,caddeo_emergent_2022,Tsaloukidis:2023bvz,Afxonidis:2023pdq}.

The hydrodynamic behavior of conventional $U(1)$ superfluids is most succinctly captured by the two-fluid model of superfluidity proposed by Laszlo Tisza and Lev Landau in the 1940s as an attempt to explain the behavior of liquid helium below its lambda point \cite{Tisza,Landau_Theory}. In this two-fluid model, superfluid helium is described as a mixture of two fluid components: the normal component and the superfluid component. The normal component behaves like an ordinary fluid; it has a well-defined viscosity and can carry heat. The fraction of the normal component decreases as the temperature is lowered and becomes zero at absolute zero temperature. The superfluid component behaves quite differently. It moves without friction, exhibiting zero entropy and zero viscosity. It cannot transport heat but can move through tiny capillaries where the normal component cannot. The two-fluid model has been successful in describing many peculiar properties of superfluid helium, and it laid the foundation for later developments in the theory of quantum liquids.

Fracton superfluids were first introduced in \cite{Chen_fractonic_2020}. Subsequently various aspects of such superfluid states have been investigated in condensed matter systems \cite{chen_fractonic_2021,PhysRevResearch.4.023151,PhysRevB.106.064511,PhysRevB.107.195132,PhysRevB.107.195131}, diffusive systems without momentum conservation \cite{Stahl:2023prt}, large $N$ theories \cite{Jensen:2022iww} and in the context of curved backgrounds \cite{Armas:2023ouk,Jain:2023nbf}. In translationally invariant fluids, at finite velocity, the dipole symmetry will in general be broken by the thermal state \cite{Glorioso_2023,Jain:2023nbf,Armas:2023ouk}, and a  dipole-associated  Goldstone  vector field $\psi_i$ is required in the long-wavelength description, signalling the spontaneous breaking of the dipole symmetry. However, it turns out that the dipole Goldstone is not independent from momentum density, at least to the leading order in derivatives\footnote{This situation is analogous to the breaking of the boost symmetry in boost-invariant fluids wherein the boost Goldstone is not independent from the velocity field \cite{Alberte:2020eil,Komargodski:2021zzy}. See also \cite{Armas:2023dcz} for an interesting case of Carrollian fluids where the Carroll boost Goldstone constitute an independent degree of freedom that cannot be eliminated from the hydrodynamic theory.}. In alignment with the terminology used in references \cite{Jensen:2022iww,Armas:2023ouk,Jain:2023nbf}, we dub this dipole condensate phase as \textit{$p$-wave fracton superfluids}. A hydrodynamic theory for this dipole-condensed phase has been investigated in a number of works \cite{PhysRevResearch.3.043186,Grosvenor:2021hkn,GloriosoLucas22,PhysRevE.107.034142,Armas:2023ouk,Jain:2023nbf}.

Nonetheless, a second symmetry breaking pattern is allowed in dipole conserving systems, corresponding to the condensation of the $U(1)$ charge, this case is dubbed the \textit{$s$-wave fracton superfluid} phase. The $s$-wave phase is characterized by both the $U(1)$ and dipole charges being spontaneously broken.
So, in spite of a new scalar  $\theta$  entering into the low energy theory, the model contains a single massless mode \cite{Pena-Benitez:2023aat}. The hydrodynamic description of this phase has been studied at the ideal level in \cite{Armas:2023ouk} and some preliminary analysis has been also performed in \cite{Jain:2023nbf}. In particular, the authors of  \cite{Jain:2023nbf, Armas:2023ouk} have identified two distinct schemes for organizing the derivative expansion and have argued that both appear consistent. Having examined both expansion schemes in detail, however, it appears to us that neither choice is in fact fully consistent. We attribute this issue to the absence of a common scaling symmetry, which prevents the assignment of a definite weight to time derivatives. In this work, we propose a gradient expansion scheme that is agnostic of this issue and construct the dissipative (linearized) third order hydrodynamic theory for $s$-wave dipole conserving superfluids. We carefully discuss the consistency of our derivative expansion and argue that it systematically organizes the hydrodynamics of fracton superfluids in the $s$-wave phase.

Our manuscript is organized as follows: Firstly, in Sec. \ref{sec:1}, we present a hydrodynamic framework for (sub)diffusion of dipole-conserving fluids with an internal dipole degree of freedom. We analyze the linear response, describing the phenomena of dipole relaxation, and discuss various physical regimes in which dipole density significantly influences the low-energy dynamics. 
Next, in Sec. \ref{sec:2}, we construct a Landau-Ginzburg model containing a Goldstone vector field transforming non-linearly under the dipole symmetry and a scalar field transforming linearly under $U(1)$ transformation. The scalar field is understood as the $U(1)$ order parameter. Therefore,  to model the $U(1)$ spontaneous symmetry breaking we add to the Lagrangian a Mexican hat potential for the order parameter. The low-energy degrees of freedom of the broken-phase are discussed.
In Sec. \ref{sec:3} we construct the ideal superfluid theory using the Hamiltonian formalism. Then in  Sec. \ref{sec:4}, we extend the ideal theory to a dissipative regime. We focused in the hydrodynamic sector of the theory containing only gappless modes. To conclude in Sec. \ref{sec:discussion} we discuss our results and comment on some outlooks.

\subsection{Summary of the results}

\begin{itemize}
    \item \textbf{Hydrodynamic model for (sub)diffusion of fractons}:
In a dipole-conserving system with an intrinsic dipole moment there is an additional continuity equation corresponding to the non-conservation of an internal dipole:
\begin{equation}
    \partial_\mu K^{\mu i}  = - J^i\,,
\end{equation}
where $K^{\mu i}$ and $J^\mu$ are the intrinsic dipole and $U(1)$ currents, respectively. The theory has one hydrodynamic (longitudinal) degree of freedom with dispersion relation
\begin{equation}
    \omega = -iD(k)k^2\,,
\end{equation}
the diffusion coefficient takes the form $D\sim k^2$. In addition, the theory has two gapped modes, one longitudinal and the other transverse. For systems with a gap taking macroscopic values ($\Gamma\sim\omega\sim k^2$), these modes become quasi-hydrodynamic with a dispersion relation
\begin{equation}
    \omega\sim -i(\Gamma + k^2)\,.
\end{equation}
    
    \item \textbf{$s$-wave fractonic superfluid}:
The superfluid phase is characterized by $d+1$ fields transforming non-linearly: a scalar $\theta$ and a vector  $\psi_i$, respectively. However, the non-trivial commutation relation between space translations and dipole transformations (see Eq. \eqref{eq:commutations}) implies that only one excitation will be gapless ($\theta$), whereas the combination $v_i^s=\partial_i\theta-\psi_i$ will be massive, with a mass proportional to the expectation value of the $U(1)$ order parameter $v$. We dubbed the field $v_i^s$ as invariant superfluid velocity; therefore, for small enough $v$, the mass of $v_i^s$ could be of the order of the macroscopic length and time scales, and the field could become quasi-hydrodynamic.
  
The degrees of freedom in the hydrodynamic regime are the energy $\epsilon$, charge $n$, momentum of the thermal non-fractonic excitations $\tilde p_i$, and the scalar Goldstone $\theta$. The system has two sound modes:
\begin{equation}
    \omega_\alpha \sim \pm c_\alpha(k) k - i D_\alpha(k)k^2\,,
\end{equation}
where $c_1(k)\sim c+k^2$, $D_1(k)\sim \Gamma+k^2$, and $c_2(k)\sim k+k^3$, $D_2(k)\sim \Omega$. Contrary to ordinary superfluids, the speed of sound of the second mode vanishes linearly with $k$ when $k\to 0$. In addition, there is a shear (transverse) mode:

\begin{equation}
    \omega \sim -i\, k^2\,\,.
\end{equation}
On the other hand, the intrinsic dipole density $K^{0i}$ and the scalar superfluid velocity $v_i^s$ are gapped and may be relevant only in the quasi-hydrodynamic regime. They add two extra longitudinal and transverse modes to the previous hydrodynamic modes, respectively. Although we do not explicitly study this regime associated with the dipole Goldstone, we expect extra longitudinal and transverse modes with gaps:

\begin{equation}
    \omega_\alpha\sim \pm\mathcal M_\alpha -i\,\Gamma_\alpha\,,
\end{equation}

where $\mathcal M_\alpha,\Gamma_\alpha\geq 0$. Associated with the intrinsic dipole density, there will also be longitudinal and transverse modes with dispersion relations:

\begin{equation}
    \omega_\alpha\sim -i\,\tilde\Gamma_\alpha\,.
\end{equation}

Notice that at the leading order in $k$, these two sets of modes are distinguishable as long as $\mathcal M_\alpha\neq 0$.

Our multi-component general description includes a normal component with momentum $\tilde p_i$ and two ``superfluid" components with velocities $u_i=\partial_i\theta$, $\xi_{ij}=\partial_i\psi_j$. However, only the thermal momentum and the dipole superfluid velocity $(\xi_{ij})$ are thermodynamic variables. Therefore, in the hydrodynamic regime, the superfluid is effectively a two-component fluid.

\end{itemize}

\section{(Sub)diffusion with intrinsic dipole}\label{sec:1}
Hydrodynamics is generically understood as the late time effective description of systems near thermal equilibrium. In the regime of its validity, the \textit{hydrodynamic regime}, a hydrodynamic theory is determined by the gapless excitations corresponding to the conserved quantities. All non-conserved quantities are then assumed to have already relaxed onto a local equilibrium state. However, in certain physically relevant cases it may be viable to incooperate degrees of freedom that are not strictly hydrodynamic and posses a finite frequency part in the zero wavevector limit. Such excitations have a finite relxation time $\tau$ and therefore can be neglected at long enough times $t\gg\tau$ but are relevant at intermediate times $t\sim\tau$. This approach is justified whenever the relaxation time is made small in the perturbative sense. For example, in systems exhibiting weak explicit breaking of the underlying symmetries wherein the non-hydrodynamic modes correspond to quantities that are almost conserved \cite{Burgess:1998ku,PhysRevB.96.195128,PhysRevD.99.086012,PhysRevLett.128.141601,PhysRevD.108.086011,PhysRevD.108.L081903}. Similarly, in the study of spinful fluids one introduces a non-hydrodynamic degree of freedom associated with the spin density \cite{lukaszewicz_micropolar_1999,Gallegos:2021bzp,Hongo:2021ona,Gallegos:2022jow}.

In this section, we will consider an effective theory, which captures the dynamics of fluids with intrinsic dipole moments. Our construction introduces a non-hydrodynamic degree of freedom, which may be relevant depending on the relaxation time-scale. 
\subsection{Conservation laws}
        In a dipole conserving system it can be shown that generically the conserved charges satisfy the following set of continuity equations
\begin{align}
\label{eq_moncons}  \partial_\mu J^\mu  = 0\,,\\
\label{eq_todipcons}  \partial_\mu J^{\mu i}  = 0\,,
\end{align}
where dipole current density $J^{\mu i}=\mathcal J^{\mu i}+K^{\mu i}$ split into two contributions, the first one dub orbital dipole current $\mathcal J^{\mu i}=\big(nx^i,J^jx^i\big)$, whereas the second contribution is called intrinsic dipole current $ K^{\mu i}=\big(\pi^i,K^{ji}\big)$. After combining Eqs. \eqref{eq_moncons} and \eqref{eq_todipcons}, the conservation equations can be recast in the form 
\begin{align}
\label{eq_moncons1}  \partial_\mu J^\mu  = 0\,,\\
\label{eq_dipcons}  \partial_\mu K^{\mu i}  = - J^i\,.
\end{align}

Given this set of conservation equations the monopole-dipole symmetry can be gauged introducing the set of gauge fields $(A_\mu,B_{\mu i})$ with transformation rule
\begin{equation}
    \delta A_\mu = \partial_\mu\alpha - \beta_i\delta^i_\mu\,,\qquad \delta B_{\mu i} = \partial_\mu\beta_i\,.
\end{equation}
With these gauge transformations the minimally coupled fractonic theory
\be \label{eq:fractonTheory}
S = S_{\text{fractons}}+\int d^{d+1} x \Big(  J^{\mu} A_{\mu} + K^{\mu i} B_{\mu i} \Big) 
\ee
will obey the conservation equations Eqs. \eqref{eq_moncons1}, and \eqref{eq_dipcons} once gauge invariance is requested.

Actually,  let us to emphasize that the dipole gauge field does not need to be symmetric at this point $B_{ij} \neq B_{ji}$. Theory \eqref{eq:fractonTheory} admits a hydrostatic equilibrium (see \cite{Jensen:2013kka} for an introduction to hydrostatics) specified by the set of isometry parameters $\mathcal K = \{\Lambda, \Sigma_i\}$ such that 
\be \label{eq:isometry}
\delta_{\mathcal K } A_{\mu}= \partial_\mu\Lambda - \Sigma_i \delta^i_\mu = 0 \,, \qquad \delta_{\mathcal K } B_{\mu i} = \partial_\mu\Sigma_i = 0\,.
\ee 
In order to make contact with the conventional treatment of hydrodynamics, we find it convenient to identify the parameters $\mathcal K = \{\Lambda, \Sigma_i\}$ with the local chemical potentials for charge $\mu=\Lambda$ and dipole $\mu_i=\Sigma_i$. Then, the isometry conditions, corresponding to the hydrostatic configurations, can be expressed as follows 
 \be \label{eq:hydrostatic}
\partial_t \mu =0\,, \quad \partial_i \mu - \mu_i = 0\,, \quad \partial_t \mu_i = 0\,, \quad \partial_i \mu_j = 0\,. 
 \ee 
 Therefore, the most general equilibrium configuration can be parameterized in terms of the two constant parameters $\mu^0$ and $\mu^0_i$ with $\mu_i=\mu^0_i$ and $\mu=\mu_0 + \mu^0_i x_i$. 

\subsection{Leading order hydrodynamics}\label{sec:2.2}
We postulate that the generalized form of the first law in the presence of an intrinsic dipole moment takes the following form\footnote{Our construction is in that respect analogous to the theory of spin hydrodynamics \cite{Gallegos:2021bzp,Hongo:2021ona,Gallegos:2022jow} wherein the first law is extended in order to account for the regime where spin remains an independent dynamical quantity.}
\be \label{eq_firstLawdiff}
T ds =  - \mu dn - \mu_i d \pi^i \,,
\ee 
where $T$ is the fixed temperature of the environment assumed to be in thermal equilibrium. Using the conservation equations and Eq. \eqref{eq_firstLawdiff} we can express the entropy production in the form
\be \label{eq:secondLawInternal}
T \partial_\mu  S^\mu =  J^i  \Big(\mu_i -\partial_i \mu\Big) - K_{ij} \partial_i  \mu_j \,,
\ee
with the entropy current defined as $S^\mu = \big(s,\, -\frac{\mu}{T}  J^i - \frac{\mu_j}{T}  K^{ij} \big)$. Imposing the second law of thermodynamics locally $T \partial_\mu  S^\mu\geq0$ then fixes the form of the constitutive relations for the hydrodynamic currents up to a desired order in a derivative expansion.

We now proceed to establish a systematic power counting scheme. From the hydrostatic conditions Eq. \eqref{eq:hydrostatic} it follows that in an equilibrium configuration $\mu_i = \partial_i\mu$. In other words, dipole chemical potential only becomes an independent degree of freedom in out-of-equilibrium processes and in particular vanishes in the state of the thermodynamic equilibrium in the absence of external sources. Therefore, we infer that $\{\mu_i\,, \pi_i \} \sim \mathcal{O}(\nabla)$. Such counting is in fact analogous to the counting of the spin chemical potential as order one quantity in the hydrodynamics of fluids with spin \cite{Gallegos:2021bzp,Hongo:2021ona,Gallegos:2022jow}. We also assume that the time derivatives are counted as $\partial_t \sim \mathcal{O}(\nabla^2)$. 
 
Implementing our derivative counting scheme, we now determine the leading order hydrodynamics corresponding to the truncation of the entropy production formula \eqref{eq:secondLawInternal} at the second order in spatial gradients. We conclude that the most general expression for the first order charge current reads
\be 
J^i_{(1)}  = \beta \Big(\mu_i - \partial_i \mu \Big)
\ee   
where $\beta \geq 0$ is a dissipative transport coefficient that will play a key role in the dynamics of the dipole sector. Plugging the constitutive relations into the equations of motion we arrive at the leading order hydrodynamic equations
\be\begin{split}
\partial_t n + \beta \partial_i \Big( \mu_i - \partial_i \mu \Big) &= \mathcal{O}(\nabla^4)\,, \\ 
\beta \Big(\mu_i - \partial_i \mu \Big) &=\mathcal{O}(\nabla^3)\,.
\end{split}
\ee 
Importantly, the dipole Ward identity does not constitute a dynamical equation at this order but rather serves as a constraint. From this constraint it follows that there are two distinct possibilities:
\be 
\beta \sim \mathcal{O}(1)\,\, \text{and}\,\, \mu_i - \partial_i \mu \sim  \mathcal{O}(\nabla^3) \quad \text{or} \quad \beta \sim \mathcal{O}(\nabla^2)\,\, \text{and}\,\, \mu_i - \partial_i \mu \sim  \mathcal{O}(\nabla)\,.
\ee 
The former condition corresponds to the \textit{pure hydrodynamic regime} where the dipole density is relaxed and does not lead to an independent degree of freedom. In this regime, the hydrostatic condition $\mu_i = \partial_i \mu$ holds dynamically in a derivative expansion \cite{Gallegos:2022jow}. In fact, we shall see that this regime is equivalent to the theory of subdifussion described in \cite{Gromov_hydro_2020}.

On the other hand, the latter possibility correspond to the \textit{dipole dynamical regime} wherein the transport coefficient is made perturbatively small $\beta \sim \mathcal{O}(\nabla^2)$ such that the dipole density is not relaxed on the hydrodynamic timescales and undergoes independent dynamics. To realize this regime one requires that the relaxation of dipole density is macroscopic such that there is a separation of scales between the relaxation of the fast modes and the former (see \cite{Hongo:2021ona,PhysRevLett.128.141601,chagnet2023hydrodynamics,PhysRevB.107.155108,PhysRevD.108.086011,gouteraux2023drude,PhysRevD.108.L081903} for analogous studies in different systems).

Focusing on the dynamics of linear fluctuations, we take $n=n_0+\delta n$, and $\pi^i=\delta\pi^i$ and expand the entropy density function up to the second order in deviations from the equilibrium state
\be 
s = s_0  - \frac{\mu_0}{T_0} \delta n  - \frac{\chi_{n}}{2T_0} \delta n^2 - \frac{\chi_{\pi}}{2T_0} \delta \pi_i^2 \,.
\ee 
Thermodynamic stability requires that
\be 
\chi_{n}\,, \chi_{\pi} \geq 0 \,,
\ee 
and the relations between the chemical potentials and densities are  
\be \begin{split} \label{eq:thermoIdentities}
 \mu &=   \mu_0 + \chi_{n} \delta n \,, \\
\mu_i &=   \chi_{\pi} \delta \pi_i\,.  
\end{split}
\ee 
In the following, we will determine the constitutive relations of the dipole and charge currents in the two regimes expanded up to the second and third order respectively\footnote{Notice that dipole equation $\partial_t \pi_i + \partial_j K^{ji} + J^i = 0$ implies that the dipole current should be expanded to one order lower that the charge currents such that both contributions have matching orders.}. We then compute the dispersion relations of the associated modes. 
\begin{table}[h!]
    \centering
   \begin{tabular}{|l|c|c|c|}
   \hline
& First order & Second order & Third order  \\
 \hline
  Vectors   & - & - &$\mu_i - \partial_i\mu$\\
  Tensors & - &$\partial_i \mu_j$&-\\
  \hline
 \end{tabular} 
    \caption{Non-hydrostatic data in the pure hydrodynamic regime. }
    \label{tab:Table1}
\end{table}
\subsection{Pure hydrodynamic regime}\label{sec:pureHydro}
In the pure hydrodynamic regime, the expressions for the currents consistent with the positivity of the entropy production \eqref{eq:secondLawInternal} are simply given by (see Table \ref{tab:Table1})
\be \begin{split}
J^i_{(3)} &= \beta \Big( \mu_i - \partial_i \mu\Big) \,, \\
\mathcal{K}^{ij}_{(2)} &= -\sigma^{ijkl} \partial_k \mu_l \,.
\end{split}
\ee 
where $\sigma^{ijkl} = \sigma_1 \delta_{ij}\delta_{kl} + \sigma_2 \delta_{i\langle k}\delta_{l \rangle j} + \sigma_3\delta_{i[k}\delta_{l]j}$ with $\beta\,, \sigma_1\,, \sigma_2\,, \sigma_3 \geq 0$. Notice that using the leading order constraint $\mu_i = \partial_i \mu + \mathcal{O}(\nabla^3)$ we can equivalently write $\mathcal{K}^{ij}_{(2)}=-\sigma^{ijkl} \partial_k \partial_l \mu$. Therefore, the hydrodynamic equations read
\be 
\begin{split}\label{eq:hydroEqsPure}
    \partial_t \delta n + \beta \partial_i \Big( \mu_i - \partial_i \mu \Big) & = \mathcal{O}(\nabla^5)\,, \\
    \partial_t \delta \pi_i + \beta \Big( \mu_i - \partial_i \mu \Big) - \sigma^{jikl} \partial_j \partial_k \partial_l \mu&=\mathcal{O}(\nabla^4)\,.
\end{split}
\ee

Using \eqref{eq:thermoIdentities} we can express the dipole equation as follows:
\be 
    \frac{1}{\chi_{\pi}}\Big(  \partial_t  + \beta \chi_{\pi}\Big) \mu_i - \beta \partial_i \mu - \sigma^{jikl} \partial_j \partial_k \partial_l \mu =\mathcal{O}(\nabla^4)\,.
\ee 
The above equation takes the form of a relaxation equation for the variable $\mu_i$ with a relaxation time $\tau = \frac{1}{\beta \chi_{\pi}}$. Since in the pure hydrodynamics regime $\beta \sim \mathcal{O}(1)$ while $\partial_t \sim \mathcal{O}(\nabla^2)$ one can safely approximate $\Big(  \partial_t  + \beta \chi_{\pi}\Big) \approx \beta \chi_{\pi}$ neglecting the evolution of the dipole density (see also \cite{10.21468/SciPostPhys.10.1.015}). After doing so, the dipole equation becomes an algebraic equation implying 
\be \label{eq:muOut}
\mu_i = \partial_i\mu + \beta^{-1}\sigma^{jikl} \partial_j \partial_k \partial_l \mu + \mathcal{O}(\nabla^4)\,.
\ee
Notice that this relation can also be expressed as $J^i_{(3)} = \partial_j \mathcal K^{ji}_{(2)}$. Understanding that $\mu_i$ will be relaxed on a ``microscopic" timescale\footnote{This is not true in the dipole dynamical regime where $\beta$ is made small in a perturbative sense.} one could directly impose $\partial_t \pi_i = 0$ in the dipole equation. Then, the constraint $J^i = - \partial_j \mathcal{K}^{ji}$ will hold to all orders leading to the generalized continuity equation for charge transport
\be  \label{eq:generalizedContinuity}
\partial_t n + \partial_i \partial_j \mathcal{J}^{ij} = 0\,, \quad \text{where} \quad \mathcal{J}^{ij} = - \mathcal{K}^{ij}\,.
\ee

Returning to the hydrodynamic equations \eqref{eq:hydroEqsPure}, the equation for charge can now be solved for the evolution of the charge density modulations after plugging \eqref{eq:muOut} into the charge equation. Indeed, after using \eqref{eq:thermoIdentities} we obtain a subdiffusive mode $\omega = - i \Gamma k^4$ with $\Gamma = \chi_n \big( \sigma_1 +\frac{d-1}{d} \sigma_2\big)$. Subdiffusive relaxation of charge density is a well-known characteristic feature of systems with dipole conservation \cite{Guardado_Sanchez_2020,Gromov_hydro_2020}. 

\begin{table}[h!]
    \centering
   \begin{tabular}{|l|c|c|c|}
   \hline
 & First order & Second order & Third order  \\
 \hline
  Vectors& $\mu_i - \partial_i\mu$   & - & $\partial^2 \big(\mu_i - \partial_i\mu\big)\,, \partial_i \partial_j \mu_j$ \\
  Tensors& - & $\partial_i \mu_j\,, \partial_i \big(\mu_j - \partial_j\mu\big)$ &-\\
  \hline
 \end{tabular} 
    \caption{Non-hydrostatic data in the dipole dynamical regime.}
    \label{tab:Table2}
\end{table}
\subsection{Dipole dynamical regime}
We now focus on the dipole dynamical regime where $\beta \sim \mathcal{O}(\nabla^2)$ and $\mu_i-\partial_i\mu\sim\mathcal{O}(\nabla)$. The classification of the non-hydrostatic data is provided in the Table \ref{tab:Table2}. As in the previous section, we will include second and third order corrections to the dipole and charge currents respectively. To do so, we must  truncate the entropy production equation \eqref{eq:secondLawInternal} at fourth order in gradients $T \partial_\mu S^{\mu} = \Delta + \mathcal{O}(\nabla^5)$ where $\Delta \geq 0$ and consider the most general form of the currents $K^{ij}_{(2)}$ and $J^i_{(3)}$ consistent with the second law. Since we are interested in the linearized regime we can set $J^i_{(3)} = \partial_j j_{(2)}^{ji}$ (see Table \ref{tab:Table2}). Then, the second law of thermodynamics can be expressed as
\be 
T \partial_\mu  S^\mu = \partial_i\Big(j_{(2)}^{ij}  \big( \mu_j - \partial_j \mu\big)\Big) +  \beta \big(\mu_i-\partial_i\mu\big)^2 -  j_{(2)}^{ij} \partial_i \big( \mu_j - \partial_j \mu\big) - K^{ij}_{(2)} \partial_i  \mu_j + \mathcal{O}(\nabla^5)\,.
\ee 
Therefore, after absorbing the total derivative term into the entropy current, the positivity of the entropy production can be guaranteed so long as
\be \label{eq:2lawC}
 -  j_{(2)}^{ij} \partial_i \big( \mu_j - \partial_j \mu\big) - K^{ij}_{(2)} \partial_i  \mu_j \geq 0
\ee 
for arbitrary configurations of fluid variables $\mu$ and $\mu_i$. The most general constitutive relations for the currents are given as
\be \begin{split}
j_{(2)}^{ij}  &=  - \alpha_{ijkl} \partial_k \big(\mu_l -\partial_l \mu\big) - \gamma_{ijkl} \partial_k \mu_l\,, \\
K^{ij}_{(2)}  &= -  \sigma_{ijkl}   \partial_k \mu_l  - \bar \gamma_{ijkl} \partial_k \big(\mu_l -\partial_l \mu\big)\,.
\end{split}
\ee 
In addition, rotational invariance fixes the transport coefficients to have the following tensor structures
\be \begin{split}
\sigma_{ijkl} &=\sigma_1 \delta_{ij}\delta_{kl} + \sigma_2 \delta_{i\langle k}\delta_{l \rangle j} + \sigma_3\delta_{i[k}\delta_{l]j}\,, \\
\alpha_{ijkl} &=\alpha_1 \delta_{ij}\delta_{kl} + \alpha_2 \delta_{i\langle k}\delta_{l \rangle j}\,, \\
\gamma_{ijkl} & =\gamma_1 \delta_{ij}\delta_{kl} + \gamma_2 \delta_{i\langle k}\delta_{l \rangle j}\,, \\
\bar \gamma_{ijkl} &=\bar\gamma_1 \delta_{ij}\delta_{kl} + \bar\gamma_2 \delta_{i\langle k}\delta_{l \rangle j}\,.
\end{split}
\ee
Then, Eq. \eqref{eq:2lawC} can be expressed in a compact matrix form:
\be
\begin{pmatrix} \partial_k \big( \mu_l - \partial_l \mu \big) & \partial_k \mu_l \end{pmatrix} \begin{pmatrix} \alpha_{ijkl} & \gamma_{ijkl} \\ \gamma_{ijkl} & \sigma_{ijkl} \end{pmatrix} \begin{pmatrix} \partial_i \big( \mu_j - \partial_j \mu\big)\\ \partial_i  \mu_j  \end{pmatrix} \geq 0\,.
\ee
In passing, we have implemented the Onsager reciprocal relations fixing $\bar \gamma_{ijkl} = \gamma_{ijkl}$. Therefore, the requirement of the local second law of thermodynamics $\partial_\mu S^\mu\geq 0$ imposes the following set of constraints on the transport coefficients: 
\be 
\sigma_1\,, \sigma_2\,,\sigma_3\,, \alpha_1\,, \alpha_2 \geq0\,, \quad \sigma_1\alpha_1\geq\gamma^2_1 \,, \quad \sigma_2\alpha_2\geq\gamma^2_2 \,.
\ee 
Now we are in the position of writing the equations of motion governing dissipative dipole hydrodynamics. To do so, we plug the constitutive relations into Eqs. \eqref{eq_moncons}, \eqref{eq_dipcons}, and make use of the thermodynamic relations Eqs.\eqref{eq:thermoIdentities} to express them in term of the densities
\be \begin{split} \label{eq:linearizedEoms1}
0 & = \partial_t \delta n + \beta \chi_\pi \partial_i \delta \pi_i - \beta \chi_n \partial^2 \delta n + \alpha \chi_n \partial^4 \delta n - \big(\alpha+\gamma\big) \chi_{\pi} \partial^2 \partial_i \delta \pi_i  \,, \\ 
0 &= (\partial_t +\beta \chi_\pi)\delta \pi_i - \beta \chi_n \partial_i \delta n  +\big(\alpha+\gamma\big)  \chi_n \partial^2 \partial_i \delta n -      
     \lambda_{||} \chi_{\pi}  \partial_i \partial_j \delta \pi_j - \lambda_{\perp} \chi_{\pi} \partial^2 \delta \pi_i  \,,
\end{split}
\ee 
where 
\be \begin{split}
\lambda_{||} &=  \frac{1}{2} \Big(  2(\sigma_1+2\gamma_1+\alpha_1)  + \frac{d-2}{d} (\sigma_2+2\gamma_2+\alpha_2) - \sigma_3 \Big)\,, \\
\lambda_{\perp} &= \frac{1}{2}\Big(\sigma_2+2\gamma_2+\alpha_2+\sigma_3\Big)\,,  \\
\alpha &= \alpha_1  + \alpha_2  \frac{d-1}{d} \,, \\
\gamma &= \gamma_1  + \gamma_2 \frac{d-1}{d}  \,, \\
\sigma &= \sigma_1  + \sigma_2 \frac{d-1}{d}  \,, \\
\lambda &= \lambda_{||} + \lambda_{\perp} = \alpha+2\gamma+\sigma\,.
\end{split}
\ee

Fourier transforming, and decomposing the dipole density as $\tilde \pi_i = \tilde\pi^{||}\hat k_i + \tilde\pi^{\perp}_i$ the linearized equations of motion take a block diagonal form 
\be
\begin{pmatrix}
    -i\omega +\beta \chi_n k^2 +\alpha \chi_n k^4 & i \beta \chi_{\pi} k + i\big(\alpha+\gamma\big) \chi_{\pi} k^3 & 0  \\
-i\beta \chi_n k - i\big(\alpha+\gamma\big)  \chi_n k^3  & -i\omega +\beta \chi_{\pi} + \lambda \chi_{\pi} k^2& 0  \\
0 & 0 & -i\omega +\beta \chi_{\pi} + \lambda_{\perp} \chi_{\pi} k^2  
\end{pmatrix} 
\begin{pmatrix}
    \tilde n  \\
 \tilde \pi^{||}  \\
 \tilde \pi^{\perp}_{i}  
\end{pmatrix} =0\,.
\ee
First, we focus on the transverse sector whose eigen-vector takes the form $\textbf{v}_{\perp} = \begin{pmatrix} 0 & 0 & 1\end{pmatrix}^T$ and eigenfrequency reads
\be \label{eq:transverseDispersion}
\omega_{\perp} = - i (\beta \chi_{\pi} + \lambda_{\perp} \chi_{\pi} k^2)\,,
\ee 
as expected, this mode is purely imaginary, and non-hydrodynamic due to the non-conservation of intrinsic dipole.

On the other hand, the longitudinal sector contains the following modes
\be \begin{gathered}
\omega^{\pm}_{||} = - \frac{i}{2} \Big(  \beta \, \chi_{\pi} +  \big(\beta \, \chi_{n}+ \lambda \, \chi_{\pi}\big)k^2 +  \alpha \, \chi_{n} k^4 +\\
\pm  \sqrt{\big( \beta \, \chi_{\pi} +  \big(\beta \, \chi_{n}+ \lambda \, \chi_{\pi}\big)k^2 +  \alpha \, \chi_{n} k^4 \big)^2 - 4  \chi_{n} \, \chi_{\pi} \left( \sigma \beta + k^2 \big( \alpha \sigma - \gamma^2 \big)  \right)  k^4} \,\Big)\,.
\end{gathered}
\ee
In the low $k$ expansion, the $\omega^{-}_{||}$ corresponds to a hydrodynamic, subdiffusive mode. Indeed, expanding in small $k$ we find that the leading contribution is subdiffusive
\be \label{eq:mode1}
\omega^{-}_{||} = -i\sigma \chi_n k^4 +\mathcal{O}(k^6)\,.
\ee 
The second longitudinal mode $\omega^{+}_{||}$ is non-hydrodynamic in the low momentum expansion 
\be \label{eq:mode2}
\omega^{+}_{||}= -i\Big( \beta \, \chi_{\pi} + \left( \beta \, \chi_{n} + \lambda \, \chi_{\pi} \right) k^2 + \left( \alpha-\sigma \right) \chi_{n} k^4\Big) +\mathcal{O}(k^6)
\ee 
and decays with a relaxation time $\tau = \frac{1}{\beta \chi_{\pi}}$ just as the transverse mode \eqref{eq:transverseDispersion}. However, it is important to note that expansions \eqref{eq:mode1} and \eqref{eq:mode2} actually lie outside of the region of validity of the dynamical dipole theory. Sending $k\rightarrow0$ while keeping $\beta$ fixed contradicts the assumption $\beta \sim \mathcal{O}(\nabla^2)$ and to study this gapless regime one should resort to the pure hydrodynamic theory Section \ref{sec:pureHydro}. 

The appropriate expansion for the dipole dynamical regime is obtained by making $k$ small while keeping the ratio $\frac{\beta}{k^2}$ fixed. Employing this expansion procedure we arrive at the following dispersion relations 
\be \begin{split}
    \omega^{-}_{||} &= -ik^4  \frac{\beta \sigma +k^2\big(\alpha \sigma-\gamma^2 \big)}{\beta+k^2\lambda}\chi_n+\mathcal{O}(k^6)\,, \\
    \omega^{+}_{||} &= -i \big( \beta+k^2\lambda \big)\chi_{\pi} -ik^2 \frac{\Big(\beta + k^2\big(\alpha+\gamma \big) \Big)^2}{\beta+k^2\lambda}\chi_n +\mathcal{O}(k^6)\,.
\end{split}
\ee

\section{Dipole sigma model}\label{sec:2}
  In this section, we construct the effective zero-temperature theory for dipole-conserving systems in the $s$-wave phase where the $U(1)$ symmetry is spontaneously broken. We begin by considering the $p$-wave phase where only dipole symmetry is broken, and subsequently break the $U(1)$ symmetry to derive the effective field theory for the phase where both charge and dipole have condensed. We then provide a hydrodynamic interpretation of the model and analyse the spectrum of the linearized theory.

\subsection{Goldstone theory and symmetry breaking}
Let us now consider an explicit example of $U(1)$ symmetry breaking in the theory describing the $p$-wave phase coupled to the monopole order parameter $\Phi$. To do so, we implement the coset construction method based on the non-linear realisation of the underlying symmetry group \cite{PhysRev.177.2239,PhysRev.177.2247}. In our case, the full set of generators consist of $\mathcal J_{ij}$, $\mathcal P_i$, $\mathcal H$, $\mathcal Q$, $\mathcal D_i$ corresponding to rotation, spatial translation, time translation, $U(1)$ shift and dipole shift, respectively. Generators satisfy the fracton algebra with the following non-vanishing commutation relations 
\begin{equation}
\begin{gathered}
     [ \mathcal J_{ij}, \mathcal P_k] =  2\delta_{k[i} \mathcal P_{j]}\,, \\
      [ \mathcal J_{ij}, \mathcal D_k] =  2\delta_{k[i} \mathcal D_{j]}\,, \\ 
[\mathcal D_{i}, \mathcal P_j] = \delta_{ij} \mathcal Q\,,\\
 [ \mathcal J_{ij},  \mathcal J_{kl}] = 2\delta_{i[k}  \mathcal J_{l]j} + 2\delta_{j[l}  \mathcal J_{k]i}\,.
\end{gathered}
\label{eq:commutations}
\end{equation}

Our goal is to construct an effective theory capturing the spontaneous breaking of the dipole generator. To do so, we shall follow the standard technique for spacetime symmetry groups \cite{Volkov:1973vd,Ivanov:1975zq}. Therefore, we start first by defining an element of the coset as:
\begin{equation}
    \Omega = e^{ix^\mu \mathcal P_\mu}e^{i\psi_i \mathcal D^i}\,,
\end{equation}
the left action of the group on the coset elements $g=e^{ia^\mu \mathcal P_\mu}e^{i\beta_i \mathcal D^i+i\alpha \mathcal Q}e^{i\Theta^{ij} \mathcal J_{ij}}$ takes the form
\begin{equation}
    g\Omega = \Omega'h[x',g,\psi]\,,
\end{equation}
where $t'^0=t^0+a^0$, $x'^i=R^i\,_j(\Theta)x^j+a^i$, and 
\begin{align}
    \psi'_i(x') &= R^i\,_j(\Theta)\psi_j(x')+\beta_i\,,\\
    h[x',g,\psi] &= e^{i\lambda(x';\alpha,a^i,\beta_i)\mathcal Q }e^{ i\Theta^{ij}\mathcal J_{ij}}\,,
\end{align}
with $\lambda(x';\alpha,a^i,\beta_i)=\alpha- a^i\beta_i + x'^i\beta_i$. 
The Maurer-Cartan (MC) form  
\be \label{eq:mc}
\omega = \Omega^{-1} d \Omega 
\ee 
can be expanded in terms of the generators as
\be 
\omega_\mu = i \big(\mathcal P_\mu + D_\mu\psi_i \mathcal D^i + A_\mu \mathcal Q \big)\,.
\ee
The coefficient of the broken generator then can be interpreted as the (dipole) covariant derivative of the Nambu-Goldstone fields $D_\mu\psi_i=\partial_\mu\psi_i$. On the other hand, the components of the Maurer-Cartan form along the unbroken generator $A_\mu=-\psi_i \delta^i_\mu$ transforms as a gauge field with $\delta A_\mu = \partial_\mu\lambda$.

Consequently, the building blocks are $ \partial_t \boldsymbol\psi$, $\nabla\cdot \boldsymbol\psi$, $\partial_{[i} \psi_{j]}$, and $\partial_{\langle i} \psi_{j\rangle}$. In addition, if we allow for the presence of a field $\Phi$ charged under the unbroken generator, the ``gauge field" $A_\mu$ allows us to define the $U(1)$ covariant derivative
\be 
\mathcal D_\mu\Phi  = \left(\partial_\mu  + i A_\mu \mathcal Q\right)\Phi \,,
\ee
notice that $\Phi'(x')=h\Phi(x')$.

In what follows, we will propose a toy model capturing the phenomenon spontaneous breaking of the $U(1)$ charge. That is possible once we interpret $\Phi$ as the order parameter of monopole charge, and add a Mexican hat potential $V(|\Phi|)=\frac{\lambda}{4}(|\Phi|^2-v^2)^2$, to the minimal Lagrangian 
\be 
\mathcal{L} = \frac{1}{2} |\partial_t \Phi|^2 +  \frac{a_2}{2}  |\partial_t\psi_i|^2 -  \frac{a_3}{2}  |\mathcal D_i \Phi|^2 - \frac{C_{ijkl}}{2}\partial_i \psi_j\partial_k \psi_l - V(|\Phi|)\,,
\ee
with 
$C_{ijkl}=c_1 \delta_{ij}\delta_{kl} + c_2 \delta_{i\langle k}\delta_{l \rangle j} + c_3 \delta_{i[k}\delta_{l]j}$.
In particular, we notice that in a symmetry broken phase where $\Phi(t,x)=(v+\eta(t,x))e^{i\theta(t,x)}$ the Lagrangian reads
\begin{align}
    \mathcal L =&  \frac{1}{2} \Big[  (\partial_t \eta)^2 - a_3 (\partial_i \eta)^2 - \lambda v^2\eta^2\Big] + \frac{v^2}{2}\left[ (\partial_t\theta)^2- \ a_3\left(\psi_i - \partial_i\theta \right)^2\right]  \\
  +& \frac{1}{2}\Big[  a_2 
  (\partial_t\psi_i)^2  - C_{ijkl}\partial_i \psi_j\partial_k \psi_l \Big]  + \mathcal L_{\text{int}}\,.
\end{align}
We thus see that the combination $v_i^s= \partial_i\theta - \psi_i  $ corresponds to a massive degree of freedom\footnote{Below we will interpret $v_i^s$ as superfluid velocity.}. If this mass is large as compared with the characteristic energy scale of the system at hand one can safely integrate out the dipole Goldstone by setting $\psi_i=\partial_i\theta$. This procedure is known as the inverse Higgs constraint and has been frequently employed in prior works on fracton superfluids \cite{hirono_effective_2022,Jensen:2022iww,PhysRevResearch.5.013101,Jain:2023nbf}. Here, however, we will refrain from integrating out the dipole Goldstone, as also discussed in \cite{Armas:2023ouk}, assuming that the mass term could be small enough to affect the effective low energy description of the system. Therefore, after integrating out the $\eta$ field, the low energy theory takes the following form 
\be \label{eq:effective}
\mathcal{L}_{\text{eff}} \equiv \mathcal{L}_{\text{eff}}(\partial_t \theta\,, \partial_t \psi_i\,, \partial_i\theta - \psi_i \,,\partial_i \psi_j )\,.
\ee
In the following, we will refer to \eqref{eq:effective} as the zero temperature $s$-wave fracton superfluid phase.

\subsection{Noether's currents}
In this section we take a closer look at the effective theory \eqref{eq:effective} and derive the associated Euler-Lagrange equations and Noether currents.

We start with the differential form of Noether's identity:
\begin{align}\label{eq:identity}
0 &= \Big[ \frac{\partial \mathcal{L}_{\text{eff}}}{\partial \psi_i}
 -\partial_\mu\left(\frac{\partial \mathcal{L}_{\text{eff}}}{\partial (\partial_\mu \psi_i)}\right)  \Big] \delta \psi_i -  \partial_\mu \frac{\partial \mathcal{L}_{\text{eff}}}{\partial (\partial_\mu \theta)}  \delta \theta
\\&+ \partial_\mu\Big[  \frac{\partial \mathcal{L}_{\text{eff}}}{\partial (\partial_\mu \psi_i)} \delta  \psi_i  +  \frac{\partial \mathcal{L}_{\text{eff}}}{\partial (\partial_\mu \theta)} \delta \theta \Big] + \partial_\mu \mathcal{L}_{\text{eff}} \delta x^\mu \,.\nonumber
\end{align}
The first line contains the Euler-Lagrange equations 
\begin{align}\label{eq:euler-lagrange}
\partial_t n + \partial_i J^i &= 0 \,, \\
\partial_t \pi_i + \partial_j K^{ji} &= - J^i
\end{align}
where we have introduced the following notation: 
\be
n = \frac{\partial \mathcal{L}_{\text{eff}}}{\partial (\partial_t \theta)}\,, \quad  J^i = \frac{\partial \mathcal{L}_{\text{eff}}}{\partial (\partial_i \theta)}\,, \quad \pi_i =  \frac{\partial \mathcal{L}_{\text{eff}}}{\partial (\partial_t \psi_i)}\,, \quad K^{ij} =  \frac{\partial \mathcal{L}_{\text{eff}}}{\partial (\partial_i \psi_j)}\,. 
\ee 
From the second line in \eqref{eq:identity} we can read off the Noether currents corresponding to the global continuous symmetries. For the shift symmetry $\delta \theta = \alpha$ and $\delta \psi_i = 0$ we find the local conservation law, which is identical to the first Euler-Lagrange equation. Therefore, we can identify $n$ and $J^i$ with the conserved charge density and current respectively. On the other hand,
the conservation of the dipole moment follows from the invariance under dipole shift $\delta \psi_i = \beta_i$ and $\delta \theta = \beta_i x^i$ leading to the following local continuity equation
\begin{align}\label{eq:dipole}
\partial_t (\pi_i + n x_i) + \partial_j \Big( K^{ji} + J^j x_i \Big) = 0\,.
\end{align}
Thus, we see that the conserved  dipole density $n_i = \pi_i + n x_i$ (current $J^{ij}=K^{ij} + J^i x_j$) receives contributions from both an ``orbital" $n x_i$ ($J^i x_j$) and ``intrinsic" part $\pi_i$ ($K^{ij}$). Actually, Eq. \eqref{eq:dipole} can be derived by combining the Euler-Lagrange equations Eqs. \eqref{eq:euler-lagrange}.

Since we are assuming translational invariance, we also find the local conservation laws for momentum and energy
\begin{align}
\partial_t p_i + \partial_j \mathcal T^{ji} = 0\,, \quad \partial_t \epsilon + \partial_i \mathcal E^i =0\,,
\end{align}
where the constitutive relations read 
\begin{align}
\label{eq_edensity}\epsilon &=  n \partial_t \theta + \pi_j \partial_t \psi_j - \mathcal{L}\,, \\
\label{eq_pdensity}p_i &= -n \partial_i \theta - \pi_j \partial_i \psi_j\,, \\
\mathcal E^i &= K_{ij} \partial_t \psi_j + J^i \partial_t \theta\,, \\
\mathcal T^{ji} &= \mathcal{L} \delta_{ij} - K_{jk} \partial_i \psi_k - J^j\partial_i \theta\,.
\end{align}
From Eq. \eqref{eq_pdensity} it is natural to interpret $\nabla\theta$ as the velocity of the $s$-wave component of the condensate, whereas $\nabla\psi_i$ can be interpreted as the velocity of the $i-$component of the $p$-wave condensate. Therefore, we define the velocities $u_i, \xi_{ij}$ as
\begin{align}
    u_i &= \partial_i\theta\,,\\
    \xi_{ij} &= \partial_i\psi_j\,.
\end{align}
Notice that the dipole superfluid velocity is invariant under $U(1)$ and dipole transformations. Therefore, all superfluid flows related by a dipole transformation have the same dipole superfluid velocity. On the contrary, the monopole superfluid velocity is boosted as $\delta u_i=\beta_i$. In some sense, a dipole transformation can be understood as a boost that commutes with time translations. Thus, the state cannot be labelled by $u_i$ but instead it must be labelled by a boost invariant velocity $v_i^s=u_i - \psi_i$.

Besides,  the stress tensor will not be symmetric for this class of systems. To understand the origin of that property we consider infinitesimal rotations $\delta x_i = \Omega_{ij}x_j$ under which the fields transform in the following way 
\begin{align}
\delta \theta &= - \partial_j \theta \Omega_{jk} x_k\,, \quad \delta \psi_i = -\partial_j \psi_i \Omega_{kl} x_k + \Omega_{ij} \psi_j \,.
\end{align}
And derive the angular momentum conservation law $\partial_{\mu} L^{\mu ij} =0$ with
\begin{align}
L^{\mu ij} &= 2\Big( x_{[i} p_{j]} + \psi_{[i} \pi_{j]},\,\,   \mathcal T^{k [j} x_{i]} +  K_{k [j} \psi_{i]} \Big)\,.
\end{align}
We thus see that there is intrinsic angular momentum carried by the dipole Goldstone 
\begin{align}
S^{\mu ij} &= 2\Big( \psi_{[i} \pi_{j]},\,\,   K_{k [j} \psi_{i]} \Big)\,,
\end{align}
satisfying 
\be
\partial_\mu S^{\mu ij}  = 2\mathcal T^{[ji]}\,.
\ee 
Then, it is possible to construct a symmetric stress tensor after adding the Belinfante–Rosenfeld improvement terms $p_i \rightarrow p_i + \Delta p_i$ and $\mathcal{T}^{ij}\rightarrow\mathcal{T}^{ij}+\Delta \mathcal{T}^{ij}$ with
\be 
\Delta p_i = \frac{1}{2} \partial_j S^{0ij}\,, \quad \Delta \mathcal T^{ij} = \frac{1}{2} \partial_k \big( S^{ijk} + S^{jik} - S^{kji} \big) - \frac{1}{2} \partial_t S^{0ji}\,.
\ee 
Thus, the improved momentum density and symmetrized stress tensor take the following form
\be 
 \begin{split}
    p_i&= -n u_i - \xi_{ij}\pi_j + \partial_j(\psi_{[i} \pi_{j]})\,, \\
 \mathcal T^{ji} &= 
 \mathcal{L}_{\text{eff}} \delta_{ij} 
- J^j v^s_i - J^{(i} \psi_{j)}  + K^{(i|k} \xi_{k|j)}   -  K^{(ij)}\xi_{kk} - K^{(i |k|} \xi_{j)k} \\
 & + \partial_k K^{(i|k|} \psi_{j)}- \partial_k K^{(ij)} \psi_k\,.
 \end{split}
\ee 
It is straightforward to check that the new momentum density $p_i$ is conserved $\partial_t p_i + \partial_j \mathcal T^{ji} = 0$, and under dipole symmetry satisfies the transformation rule 
\be \begin{split}
\delta_{\beta} p_i &= -n\beta_i   + \beta_{[i}\partial_j \pi_{j]}\,, \\
\delta_\beta \mathcal T^{ij} &= -J^{(i} \beta_{j)} + \partial_k K^{(i|k|} \beta_{j)}- \partial_k K^{(ij)} \beta_k \,.
\end{split}
\ee 
The above transformation properties are consistent with those listed in \cite{PhysRevE.107.034142} provided that the internal dipole density is integrated out such that $J^i=-\partial_j K^{ji}=\partial_j  \mathcal J^{ji}$ as discussed around the Eq. \eqref{eq:generalizedContinuity}.
\subsection{A hydrodynamic perspective}
Since we are interested in the hydrodynamics description of the system, we find convenient to switch to a Hamiltonian picture via a Legendre transform
\begin{equation}
    h(n,\pi_i,v_i^s,\xi_{ij}) = n\partial_t\theta +\pi_i\partial_t\psi_i -\mathcal L_{\text{eff}}
\end{equation}
where we have identified the canonically conjugated momenta with the charge and dipole densities. In addition, we impose on the superfluid velocities the constraints 
\be  \label{eq:velocitiesConstraint}
\epsilon^{ijk} \partial_j u_k = 0\,, \quad \epsilon^{ilk} \partial_k \xi_{lj} =0\,.
\ee 
It is then possible to interpret the Hamiltonian density as a microcanonical equation of state given as 
\be \label{eq:microcanonical}
\epsilon \equiv h(n\,, \pi_i\,, v^s_i\,, \xi_{ij})
\ee 
with an analogue of the local first law
\be 
d\epsilon = \mu dn + \mu_i d\pi_i + \lambda_i dv_i^s+F_{ij}d\xi_{ij} 
\ee 
where we have introduced the conjugate variables 
\be \label{eq:thermDefs}
\mu = \frac{\partial h}{\partial n}\,, \quad \mu_i = \frac{\partial h}{\partial \pi_i}\,, \quad \lambda_i = \frac{\partial h}{\partial v^s_i}\,, \quad F_{ij} = \frac{\partial h}{\partial \xi_{ij}}\,.
\ee 
In addition, the canonical Poisson bracket for this field theory reads 
\be \label{eq:canonical}
\{F, G\}_C  = \int d^d x  \Big[ \Big(\frac{\delta F}{\delta \theta} \frac{\delta G}{\delta n} - \frac{\delta F}{\delta n} \frac{\delta G}{\delta \theta} \Big)
+ \Big( \frac{\delta F}{\delta \psi_i} \frac{\delta G}{\delta \pi_i} - \frac{\delta F}{\delta \pi_i} \frac{\delta G}{\delta \psi_i} \Big)
\Big]\,.
\ee

Equations of motion, as well as the conservation equations, can be determined directly by computing the time evolution via the canonical Poisson bracket \eqref{eq:canonical} with the Hamiltonian $\mathcal{H}=\int d^d x h$
\be 
\partial_t \mathcal F = \{\mathcal F, \mathcal{H}\}_C \,.
\ee 
In particular, choosing $\mathcal F = \{\theta, \psi_i\}$, one obtains the equations governing the evolution of the Goldstones, the so-called Josephson relations
\be \begin{split} \label{eq:josephsons}
\partial_t \theta &= \mu\,, \\
\partial_t \psi_i &= \mu_i\,.
\end{split}
\ee 
Relations \eqref{eq:josephsons} are crucial for determining a systematic power counting scheme that organizes the hydrodynamics of finite-temperature dissipative fracton superfluids. Indeed, since $v_i^s$ corresponds to a massive degree of freedom, it vanishes in the state of thermodynamic equilibrium and, as a result, should be treated as an order-one quantity $\{v_i^s, \lambda_i\} \sim \mathcal{O}(\nabla)$. Therefore, the Goldstone fields themselves ought to be counted as $\theta \sim \mathcal O(\nabla^{-2})$ and $\psi_i \sim \mathcal O(\nabla^{-1})$. Consequently, one concludes that $\partial_t \sim \mathcal O(\nabla^2)$ such that both sides of the relations \eqref{eq:josephsons} are of the same order. Finally, notice that in this counting the dipole superfluid velocity is order-zero $\xi_{ij} \sim \mathcal O(1)$.

Equivalently, we can recast the Josephson relations \eqref{eq:josephsons} in terms of the invariant superfluid velocities 
\be\begin{split}\label{eq:superfluidVelocities}
\partial_t  v^s_i  &=  \partial_i  \mu- \mu_i        \,, \\
\partial_t \xi_{ij} &= \partial_i  \mu_j \,.
\end{split}
\ee
Throughout the remainder of this study, however, we will often find it convenient to express the Josephson relations in terms of variables $v_i^s$ and $\theta$ as follows 
\be\begin{split}
    \label{eq:josephson}
\partial_t \theta &= \mu\,, \\
\partial_t v^s_i &= \partial_i  \mu- \mu_i \,.
\end{split}
\ee
On the other hand, the evolution of the densities $n$ and $\pi_i$ is governed by the same local equations as computed in Lagrangian formalism \eqref{eq:euler-lagrange} with the identification
\be 
J^i = - \lambda_i\,, \quad K^{ij} = - F_{ij}\,.
\ee 
Since we are ultimately interested in the finite temperature regime, let us also compute the Poisson brackets corresponding to the momentum density
\be \begin{split}\label{eq:poisson}
\{ p_i(\textbf x)  \,, \theta(\textbf y) \}_C &=  \partial_i \theta \, \delta(\textbf x-\textbf y)\,, \\
\{ p_i(\textbf x) \,, \psi_j (\textbf y) \}_C &= \partial_i \psi_j \, \delta(\textbf x- \textbf y)\,, \\
\{ p_i(\textbf x) \,, n(\textbf y) \}_C &=  -n \, \partial_i \delta(\textbf x-\textbf y)\,, \\
\{ p_i(\textbf x) \,, \pi_j (\textbf y) \}_C &= - \pi_j   \, \partial_i \delta(\textbf x- \textbf y)\,.
\end{split}
\ee 
Notice that above relations are compatible with the identification of the total momentum of a system as the generator of space translations.

Finally, full set of hydrodynamic equations governing the long-time dynamics of the zero-temperature Goldstone theory are given as 
\be \begin{split}
\partial_t \delta n - \partial_i \lambda_i &= 0\,, \\
\partial_t \pi_i - \partial_j F_{ji} -\lambda_i  &= 0\,,\\
\partial_t  v^s_i  +\mu_i    - \partial_i \mu    &= 0 \,, \\
\partial_t \xi_{ij} - \partial_i \mu_j &= 0\,.
\end{split}
\ee 
\subsection{Linear fluctuations}
In order to study the hydrodynamic modes of the theory, we consider small perturbations around a homogeneous fluid configuration with $\lambda_i=0$, $F_{ij}=0$ and $\mu_i =0$. This regime corresponds to fracton superfluids at rest and zero dipole chemical potential.

Expanding the Hamiltonian density around a homogeneous equilibrium up to the second order in perturbations 
\be \label{eq:hamiltonian}
h = h_0 + \mu_0 \delta n + \frac{1}{2} \chi_n \delta n^2 + \frac{1}{2} \chi_\pi \delta \pi_i^2 + \frac{1}{2} \chi_v ( v^s_i)^2  + \frac{1}{2} \lambda_1 \xi^2 + \frac{1}{2} \lambda_2  \xi_{\langle i j \rangle}^2 +\frac{1}{2} \lambda_3 \xi_{[ i j ]}^2 
\ee 
with $\delta n\,, \delta \pi_i\,, v^s_i$ and $\xi_{ij}$ regarded as small. The condition for the stability of a stationary solution implies that all susceptibilities are non-negative 
\be \label{eq:stability}
\chi_n\,, \chi_\pi\,, \chi_v\,, \lambda_1\,, \lambda_2\,, \lambda_3 \geq 0\,.
\ee
Using definitions \eqref{eq:thermDefs} we find the following identities
\be \begin{gathered}
\mu = \mu_0 + \chi_n \delta n\,, \quad \mu_i = \chi_\pi \delta \pi_i\,, \\
\lambda_i = \chi_v v^s_i\,, \quad F_{ij} =   \lambda_1 \xi \delta_{ij} +  \lambda_2  \xi_{\langle i j \rangle} + \lambda_3 \xi_{[ i j ]}\,.
\end{gathered}
\ee 
Substituting these into the equations of motion \eqref{eq:euler-lagrange} and \eqref{eq:superfluidVelocities} we obtain a set of linear differential equations\footnote{It may appear to the reader that there are more unknowns than equations. However, the number of independent degrees of freedom is reduced by conditions \eqref{eq:velocitiesConstraint} yielding a solvable set of equations. Alternatively, one can simply decompose $\xi_{ij}=\partial_j \psi_i$ and work directly with the dipole Goldstone. This is the choice that we adopt in the rest of the manuscript.}
\be \begin{split}
\partial_t \delta n - \chi_v \partial_i v^s_i &= 0\,, \\
\partial_t \pi_i - \lambda_{1} \partial_i \xi - \lambda_{2} \partial_j \xi_{\langle j i \rangle} - \lambda_{3} \partial_j \xi_{[ j i ]} - \chi_v v^s_i &= 0\,,\\
\partial_t  v^s_i  - \chi_n \partial_i \delta n + \chi_\pi \delta  \pi_i   &= 0 \,, \\
\partial_t \xi_{ij} - \chi_\pi \partial_i \delta \pi_j &= 0\,.
\end{split}
\ee 
After performing a Fourier transformation and decomposing the variables into their parallel and perpendicular components: $\tilde \psi_i = \tilde \psi_{||} \hat k_i + \boldsymbol{\tilde \psi_{\perp}}$, $\tilde \pi_i = \tilde \pi_{||} \hat k_i + \boldsymbol{\tilde \pi_{\perp}}$ and $\tilde  v^s_i = \tilde v^s_{||} \hat k_i + \boldsymbol{\tilde v^s_{\perp}}$, the linearized equations of motion can be elegantly expressed in a block diagonal matrix form
\be
\begin{pmatrix}
\mathcal M^{4 \times 4}_{||}  & \mbox{\Large 0} \\ 
\mbox{\Large 0}  & \mathcal M^{3 \times 3}_{\perp} \\
\end{pmatrix} \begin{pmatrix}
\textbf{v}_{||}  \\ 
 \textbf{v}_{\perp}   \\
\end{pmatrix} = 0
\ee 
where
\be\begin{split}
\mathcal M^{3 \times 3}_{\perp} &= \begin{pmatrix}
    - i \omega & -\chi_v & k^2 \lambda_{\perp} \\
\chi_\pi & -i \omega & 0 \\
-\chi_\pi & 0 & -i \omega
\end{pmatrix}\,, \quad \textbf{v}_{\perp} = \begin{pmatrix}
\boldsymbol{\tilde \pi_{\perp}}\, &
\boldsymbol{\tilde v^s_{\perp}}\, &
\boldsymbol{\tilde \psi_{\perp}}
\end{pmatrix}^{\intercal} \,, \\
\mathcal M^{4 \times 4}_{||} &= \begin{pmatrix}
- i \omega & 0 & -i \chi_v k & 0 \\
0 & -i \omega & -\chi_v & k^2 \lambda \\
- i k \chi_n & \chi_\pi & -i \omega & 0 \\
0 & -\chi_\pi & 0 & -i \omega
\end{pmatrix}\,, \quad \textbf{v}_{||} = \begin{pmatrix}
\tilde n\, &
\tilde \pi_{||}\, &
\tilde v^s_{||}\, &
\tilde \psi_{||}
\end{pmatrix}^{\intercal}
\end{split}
\ee
correspond to transverse and longitudinal sectors respectively and we have defined $\lambda = \lambda_1 + \lambda_2 \frac{d-1}{d}$ and $\lambda_{\perp} =  \frac{\lambda_2 + \lambda_3}{2}$. In the transverse sector we find a solution at zero $\omega_{\perp}(k)=0$, as well as a pair of gapped propagating modes 
\be\label{eq:transDisp}
\omega_{\perp}(k) = \pm \sqrt{\chi_{\pi}\chi_v + \lambda_{\perp} \chi_{\pi} k^2}\,.
\ee 
Interestingly, we observe that the system posses an intrinsic lengthscale, which we quantify by introducing the \textit{transverse dipole wavevector}
\be 
k^{\perp}_0 = \sqrt{\frac{\chi_v}{\lambda_{\perp} }}   \,.
\ee 
Depending on the ratio $\kappa = \frac{k}{k^{\perp}_0}$, the transverse modes exhibit the following asymptotic dependence on the momentum wave vector: 
\be
\omega_{\perp}(\kappa) \approx 
     \begin{cases}
        \pm m_0 (1 + \frac{\kappa^2}{2}) &\quad\text{for } \kappa\ll 1\\
       \pm m_0 \kappa  &\quad\text{for } \kappa \gg 1 \\ 
     \end{cases}
\ee
Here, we have introduced a mass term $m_0= \sqrt{\chi_{\pi}\chi_v}$. Therefore, transverse perturbations for which $\kappa \ll 1$ correspond to massive modes with a quadratic dispersion whereas in the regime when $\kappa \gg 1$ the perturbations travel with a soundlike dispersion.

For the longitudinal sector, we find four modes propagating with the following dispersion relations
\be \label{eq:disLong}
\omega^{\pm}_{||} = \pm \frac{\sqrt{ m^2_0 + \Lambda k^2 \pm \sqrt{ \left(m^2_0 + \Lambda k^2\right)^2 -4 m^2_0 \lambda \chi_n k^4  }}}{\sqrt{2}}
\ee 
where the superscript in $\omega^{\pm}_{||}$ refers to the second $(\pm)$ sign and $\Lambda = \lambda \chi_{\pi} + \chi_n \chi_v$. 
Here, as well, we identify a characteristic lengthscale that we dub \textit{longitudinal dipole wave vector}
\be 
k^{||}_{0} = \frac{m_0}{\sqrt{\Lambda}}\,,
\ee 
and examine the behaviour of the dispersion relations as a function of the ratio $\kappa = \frac{k} {k^{||}_{0}}$. In terms of $\kappa$  \eqref{eq:disLong} takes the simple form
\be 
\omega^{\pm}_{||} = \pm m_0 \frac{\sqrt{ 1 + \kappa^2 \pm \sqrt{ \left(1 + \kappa^2 \right)^2 -4 a  \kappa^4  }}}{\sqrt{2}}
\ee 
where $a=\frac{\lambda \chi_n m^2_0}{\Lambda^2} \leq \frac{1}{4}$ is a free parameter. In the small $\kappa$ expansion, the pair of modes associated to $\omega^{-}_{||}$ displays a magnonlike dispersion relation\footnote{Magnonlike propagation is a characteristic feature of dipole-conserving fluids \cite{PhysRevResearch.3.043186,PhysRevE.107.034142,GloriosoLucas22}.}, while the second pair corresponds to massive excitations that also propagate with quadratic momentum dependence
\be \begin{split}
\omega^{-}_{||} &= \pm m_0 \sqrt{a} \kappa^2 + \mathcal{O}(\kappa^4)\,, \quad \text{for } \kappa \ll 1, \\
\omega^{+}_{||} &= \pm m_0  \Big( 1 + \frac{\kappa^2}{2}\Big) + \mathcal{O}(\kappa^4) \,, \quad \text{for } \kappa \ll 1 \,.
\end{split}
\ee   
In contrast, in the large $\kappa$ regime we find two pairs of soundlike modes propagating with different velocities 
\be \
\omega^{\pm}_{||} = \pm m_0 \frac{\sqrt{1\pm\sqrt{1-4a}}}{2} \kappa\,, \quad \text{for } \kappa \gg 1\,.
\ee 
We thus see that in the regime where the perturbations have a wavelength smaller than the macroscopic dipole lengthscale $k_0^{-1}=\frac{\sqrt{\Lambda}}{m_0}$, there are two distinct linearly propagating modes propagating with different velocities.
Finally, let us point out that all modes are purely real, which follows from the condition for thermodynamic stability \eqref{eq:stability}.

\section{Ideal fracton superfluids}\label{sec:3}
In this section, we study the ideal fracton superfluids at finite temperature. In particular, we focus on translationally invariant, isotropic fluids with the $U(1)$ charge and dipole symmetries broken spontaneously\footnote{Notice that breaking the U(1) symmetry in a system with dipole conservation necessarily implies that dipole symmetry is broken as well.}. Such phases are sometimes referred to in the literature as $s$-wave fracton superfluids \cite{Jensen:2022iww,Jain:2023nbf,Armas:2023ouk}. We derive the hydrodynamic constitutive relations applying the method of Poisson brackets. We then study the spectrum of the theory and confirm the existence of three distinct types of propagating modes among which one is massive.

\subsection{Three-components fluid}
The full set of hydrodynamic equations of motion consists of the local conservation laws\footnote{Strictly speaking, internal dipole is not a conserved quantity.} supplemented with the Josephson relations, which capture the dynamics of the Goldstone fields:
\be 
\bg \label{eq:Idealeoms}
\partial_t n + \partial_i J^i = 0\,, \quad \partial_t \pi_i + \partial_j K^{ji} = - J^i\,, \\
\partial_t p_i + \partial_j T^{ji} = 0\,, \quad  \partial_t \epsilon + \partial_i E^i = 0\,, \\
\partial_t  v^s_i  =  \partial_i  \mu -  \mu_i   \,, \quad
\partial_t \xi_{ij} = \partial_i \mu_j\,.
\eg
\ee
Notice that the Josephson's relations are \textit{a priori} arbitrary, but we postulate that in the absence of dissipation, these are identical to the zero-temperature theory \eqref{eq:josephson}. We will soon confirm this ansatz using Poisson brackets.

With the equations of motion at hand, one is faced with the problem of identifying the relevant degrees of freedom known as hydrodynamic variables. Typically, the hydrodynamic variables are associated with the conserved densities and Goldstone modes if there are any. However, dipole-conserving systems are known not to be compatible with the simple identification of momentum as a hydrodynamic degree of freedom due to the tension between the dipole transformations and translational symmetry \cite{PhysRevResearch.3.043186,GloriosoLucas22,PhysRevE.107.034142}. In particular, it is only the exactly invariant variables that can enter into the Hamiltonian as low-energy variables. 

Keeping this in mind, we postulate that a finite temperature generalization of the theory Eq. \eqref{eq:microcanonical} is given by the following Hamiltonian density
\be
h \equiv h[n\,, s\,, \pi_i\,, v^s_i\,, \xi_{ij}\,, \Tilde{p}_i ]
\ee
where $v^s_i = \partial_i \theta - \psi_i$ and $\xi_{ij} = \partial_i \psi_j$ are the superfluid velocities familiar from the zero temperature theory and $\tilde{p}_i = p_i + n\partial_i\theta + \pi_j\partial_i\psi_j$ is what we dub as \textit{thermal momentum}. It is an invariant combination involving momentum that can be understood as the momentum carried by the thermal excitations. Thus, we can interpret the finite temperature $s$-wave superfluid as a three-fluid model involving two superfluid and one normal components respectively that can flow independently of one another. However, as we will see, the $U(1)$ superfluid velocity will be gapped, rendering it irrelevant in the hydrodynamic regime.

Let us also introduce a set of the conjugate variables 
\be \begin{split}\label{eq:definitions}
\mu &=  \frac{\delta h}{\delta n}\,, \quad \mu_i =  \frac{\delta h}{\delta \pi_i}\,, \\ 
T &=   \frac{\delta h}{\delta s}\,, \quad 
\Tilde{V}_i = \frac{\delta h}{\delta \Tilde{p}_i}\,, \\
\lambda_i &= \frac{\delta h}{\delta v^s_i}\,, \quad F_{ij} = \frac{\delta h}{\delta \xi_{ij}}\,,
\end{split}
\ee 
such that the differential of the Hamiltonian density is given as 
\be \label{eq:firstLaw}
dh = \mu dn + \mu_i d \pi_i + T ds + \Tilde{V}_i d \Tilde p_i + \lambda_i d v^s_i + F_{ij} d \xi_{ij}\,.
\ee 
The above relation can be understood at the first law of thermodynamics for the $s$-wave fracton superfluids.

\subsection{Poisson bracket method}
A simple way to obtain the ideal form of the constitutive relations in hydrodynamics is the Poisson bracket method. Once the commutators between the hydrodynamic variables are known and the Poisson bracket structure is established, the equations of motion for any function of these variables may be determined directly via 
\be \label{eq:evolution}
\partial_t \mathcal{A} = \{\mathcal{A}\,, \mathcal{H} \}\,.
\ee 
Furthermore, the structure of the Poisson bracket is universal, determined by the generic physical considerations such as symmetries rather than the microscopic details of a particular model. In fact, Poisson brackets were successfully applied in order to derive the hydrodynamic constitutive relations for the ideal fracton fluids \cite{PhysRevResearch.3.043186}. Here we combine the developments of \cite{PhysRevResearch.3.043186} with previous studies of conventional superfluids \cite{son_hydrodynamics_2001,PhysRevD.77.025004}.

For the $s$-wave superfluid, all non-trivial commutation relations between the hydrodynamic variables may be obtained in the zero-temperature theory using the canonical Poisson bracket \eqref{eq:canonical}. Then, it is not too hard to infer that the appropriate noncanonical Poisson bracket structure capturing all commutation relations is given by
\be \begin{split}\label{eq:noncanonical}
\{F, G\}_{NC}  = &- \int d^d x  \Big[ p_i \Big(\frac{\delta F}{\delta p_j}  \partial_j \frac{\delta G}{\delta p_i} - \frac{\delta G}{\delta p_j}  \partial_j \frac{\delta F}{\delta p_i}\Big) + s \Big(\frac{\delta F}{\delta p_j}  \partial_j \frac{\delta G}{\delta s} - \frac{\delta G}{\delta p_j}  \partial_j \frac{\delta F}{\delta s}\Big) \\
&+ n \Big(\frac{\delta F}{\delta p_j}  \partial_j \frac{\delta G}{\delta n} - \frac{\delta G}{\delta p_j}  \partial_j \frac{\delta F}{\delta n}\Big) + \pi_i \Big(\frac{\delta F}{\delta p_j}  \partial_j \frac{\delta G}{\delta \pi_i} - \frac{\delta G}{\delta p_j}  \partial_j \frac{\delta F}{\delta \pi_i}\Big) \\
&-\partial_i \theta \Big(\frac{\delta F}{\delta p_i} \frac{\delta G}{\delta \theta} - \frac{\delta G}{\delta p_i} \frac{\delta F}{\delta \theta} \Big) -\partial_i \psi_j \Big(\frac{\delta F}{\delta p_i} \frac{\delta G}{\delta \psi_j} - \frac{\delta G}{\delta p_i} \frac{\delta F}{\delta \psi_j} \Big)\\
&-\Big(\frac{\delta F}{\delta \theta} \frac{\delta G}{\delta n} - \frac{\delta G}{\delta \theta} \frac{\delta F}{\delta n} \Big)- \Big( \frac{\delta F}{\delta \psi_i} \frac{\delta G}{\delta \pi_i} - \frac{\delta G}{\delta \psi_i} \frac{\delta F}{\delta \pi_i} \Big)
\Big]\,.
\end{split}
\ee 
The first four terms in the equation arise from the identification of momentum as a generator of translations, while the next two lines follow directly from \eqref{eq:poisson}. The final two lines, on the other hand, are a consequence of the fact that the charge and dipole Goldstone modes and their corresponding charges are canonically conjugated \eqref{eq:canonical}, which is a direct generalization of the $U(1)$ case (see e.g. \cite{staruszkiewicz_quantum_1989}).

Using the noncanonical Poisson structure, as well as the definitions \eqref{eq:definitions}, it is then a straightforward exercise to determine the explicit form of the equations of motion by computing the evolution of the hydrodynamic variables via \eqref{eq:evolution}. From the equations of motion corresponding to the hydrodynamic densities we may read off the constitutive relations 
\be \begin{split}\label{eq:ideal}
J^i &= - \lambda_i \,, \\
K^{ij} &= - F_{ij}\,, \\
S^i &= s \Tilde V_i\,, \\
T^{ji}&=P \delta_{ij} + F_{jk}  \xi_{ik} + \lambda_j \partial_i \theta + \Tilde{V}_j \Tilde{p}_i\,.
\end{split}
\ee 
where the pressure has been defined as
\be \label{eq:pressure}
P  = n \mu + \pi_i \mu_i + Ts + \Tilde{V}_i \Tilde{p}_i - h \,.
\ee
Interestingly, we notice that the fractonic nature of the superfluid is manifested in the fact that the $U(1)$ currents receives contribution only from the superfluid velocity whereas both the superfluid velocities and normal velocity contribute to the momentum flow. In fact, if  we compute the Poisson bracket between the thermal momentum $\tilde p_i$ and the charge densities we obtain
\begin{equation}\label{eq_neutralityP}
    \{\tilde p_i (\mathbf{x}), n(\mathbf{y})\} =   \{\tilde p_i (\mathbf{x}), \pi_j(\mathbf{y})\} = 0 \,,
\end{equation}
which suggests that the thermal momentum is fracton charge ``neutral", and what in the ordinary superfluid model  would correspond to the normal component, in this case is made of thermal excitations that do not carry neither $U(1)$ nor dipole charges.

On the other hand, equations for the Goldstones lead to the same form of the Josephson's relations as derived in \eqref{eq:josephson}. In addition, it is possible to symmetrize the stress tensor by the addition of the suitable improvement terms. The improved momentum density reads 
\be 
p_i \rightarrow p_i +\frac{1}{2}\Big( \pi_j \partial_j \psi_i + \psi_i \partial_j \pi_j - \pi_i 
 \partial_j \psi_j  - \psi_j
 \partial_j \pi_i \Big)
\ee 
while the symmetric stress tensor is 
\be \begin{split}
T^{ij} &= P \delta_{ij}+\Tilde{V}_i \Tilde{p}_j  +\lambda_i v^s_j + \lambda_{(i} \psi_{j)}  - F^{(i|k} \partial_{k|} \psi_{j)}   +  F^{(ij)} \partial_k \psi_k + F^{(i |k|} \partial_{j)} \psi_k  \\
& - \partial_k F^{(i|k|} \psi_{j)}+ \partial_k F^{(ij)} \psi_k\,.
\end{split}
\ee 
Energy current cannot be directly computed in the Poisson bracket formalism, however, can be obtained using the first law of thermodynamics. Indeed, from \eqref{eq:firstLaw} after using the equations of motion \eqref{eq:Idealeoms} we find that the energy current is given as 
\be 
E^i = \Tilde{V}_i (sT + \Tilde{p}_j \Tilde{V}_j ) - \mu \lambda_i - \mu_j F_{ij}\,.
\ee 
With this we conclude our analysis of the ideal constitutive relations.
\subsection{Hydrodynamic modes}
Let us now turn our attention to the study of the hydrodynamic modes of ideal superfluids with dipole symmetry. For this purpose, we consider small deviations away from the stationary background: 
 \be 
 \bg
 n=n_0 + \delta n\,, \quad \Tilde{p}_i =  \delta\Tilde{p}_i \,, \quad \pi_i =  \delta \pi_i \,,  \\
 \epsilon= \epsilon_0 + \delta \epsilon\,, \quad \psi_i = \delta \psi_i\,, \quad  \theta =  \mu_0 t + \delta \theta\,.
 \eg
 \ee 
The dynamics of the perturbations is captured by the following set of linear equations 
 \be \begin{split}
\partial_t \delta n - \partial_i \delta \lambda_i &= 0\,, \\
\partial_t \pi_i -\partial_j \delta F_{ji} - \delta \lambda_i &= 0\,, \\
\partial_t \epsilon + s_0 T_0 \partial_i \delta \Tilde{V}_i - \mu_0 \partial_i \delta \lambda_i &= 0\,, \\
\partial_t \delta \tilde p_i + s_0 \partial_i \delta T   &=0 \,, \\
\partial_t  v^s_i  - \Big( \partial_i \delta \mu - \delta  \mu_i  \Big) &= 0 \,, \\
\partial_t \xi_{ij} + \partial_i \delta \mu_j &= 0\,.
\end{split}
\ee
To close the system and determine the evolution of the hydrodynamic modes, it is necessary to express the fluctuations in thermodynamics variables $\{\delta \mu\,, \delta \mu_i\,, \delta T\,, \Tilde V_i\,, \lambda_i\,, \delta F_{ij} \}$ in terms of the fundamental quantities $\{n\,, \pi_i\,, \epsilon\,, \Tilde p_i\,, v^s_i\,, \xi_{ij} \}$\footnote{These are the variables associated with quantities that are well-defined microscopically.}.

To achieve this, we expand the entropy density function up to the second order in perturbations
 \be\begin{split} \label{eq:entropyDensity}
s &= s_0+\frac{1}{T_0} \delta \epsilon  - \frac{\mu_0}{T_0} \delta n - \frac{1}{2 T_0}\chi_{nn} \delta n ^2 - \frac{1}{2 T_0}\chi_{\epsilon \epsilon} \delta \epsilon ^2 + \frac{1}{T_0} \chi_{n\epsilon} \delta \epsilon \delta n \\
     &- \frac{1}{2T_0}\lambda_{ijkl} \xi_{ij} \xi_{kl} - \frac{\chi_v}{2T_0}  (\delta v^s_i)^2 - \frac{\chi_p}{2T_0} \delta \Tilde{p}_i^2 - 
     \frac{\chi_\pi}{2T_0} \delta \pi_i^2 
     + \frac{\chi_{vp}}{T_0} \delta v^s_i \delta \Tilde{p}_i \,,
     \end{split}
 \ee 
where $\lambda_{ijkl} = \lambda_1 \delta_{ij} \xi + \lambda_2 \xi_{\langle i j \rangle} + \lambda_3  \xi_{[i j]}$ is the most general SO(3) invariant tensor. Using the definitions \eqref{eq:definitions} we obtain the fluctuations in the conjugate variables 
\be \begin{split} \label{eq:thermo}
 \delta T &=  T_0 \chi_{ee} \delta e - T_0 \chi_{ne} \delta n \,, \\
 \delta \mu &=  \chi_e \delta e +  \chi_n \delta n\,, \\
\delta \mu_i & =  \chi_{\pi} \delta \pi_i\,, \\
\delta \tilde V_i &=   \chi_{p} \delta \Tilde{p}_i \,,\\
\delta  \lambda_i &=    \chi_{v} \delta (\partial_i \theta - \psi_i) \,,\\
\delta F_{ij} &=   \lambda_{ijkl} \partial_k \psi_l = \lambda_1 \delta_{ij}\partial_k \psi_k + \lambda_2 \partial_{\langle i} \psi_{ j \rangle } + \lambda_3  \partial_{[ i} \psi_{ j ] } \,,
\end{split}
\ee 
where we have defined
\be
 \chi_e = \Big( \mu_0 \chi_{ee} - \chi_{ne} \Big)\,, \quad  \chi_n = \Big( \chi_{nn} - \mu_0  \chi_{ne} \Big)\,.
\ee 
Then, the set of equations can be expressed in a matrix form as 
\be
\begin{pmatrix}
\mathcal M^{6 \times 6}_{||}  & 0 \\ 
0  & \mathcal M^{3 \times 3}_{\perp} \\
\end{pmatrix} \begin{pmatrix}
\textbf{v}_{||}  \\ 
 \textbf{v}_{\perp}   \\
\end{pmatrix} = 0\,,
\ee 
where 
\be 
\textbf{v}_{||} = \begin{pmatrix}
\tilde n\, \\
\tilde \pi_{||}\, \\ \tilde \epsilon\, \\
\tilde p_{||}\, \\
\tilde \theta\, \\
\tilde \psi_{||}
\end{pmatrix}\,, \quad \mathcal M^{6 \times 6}_{||} =
\begin{pmatrix}
-i \omega & 0 & 0 & 0 & \chi_v k^2 & i \chi_v k \\
0 & -i\omega & 0 & 0 & -i\chi_v k & \chi_v + \lambda k^2 \\
0 & 0 & -i\omega & i s_0 T_0 \chi_{p} k & \mu_0 \chi_v k^2 & i\mu_0 \chi_v k \\
-i s_0 T_0\chi_{ne}  k & 0 & i s_0 T_0\chi_{ee}  k & -i\omega & 0 & 0 \\
- \chi_e & 0 & - \chi_n & 0 & -i \omega & 0 \\
0 & -\chi_{\pi} & 0 & 0 & 0 & -i\omega  \\
\end{pmatrix}
\ee 
\be
\textbf{v}_{\perp}  = \begin{pmatrix}
 \boldsymbol{\tilde\pi_{\perp}}\\
 \boldsymbol{\tilde p_{\perp}} \\
 \boldsymbol{\tilde\psi_{\perp}}
\end{pmatrix}\,, \quad \mathcal M^{3 \times 3}_{\perp} = \begin{pmatrix}
-i \omega & 0 & \chi_v + \lambda_2 k^2 \\
0 & -i \omega & 0 \\
-\chi_{\pi} & 0 & -i \omega
\end{pmatrix}
\ee 
correspond to longitudinal and transverse fluctuations respectively.

In the longitudinal sector, we find 3 pairs of modes. The first pair constitutes of the soundlike modes 

\be\label{eq:sound}
\omega = \pm v_s k + \mathcal{O}(k^2)\,, \quad v_s = 
 s_0 T_0 \sqrt{ \chi_{ee}  \chi_p }
 \,.
\ee 
For the second pair we find magnonlike dispersion 
\be\label{eq:mangon}
\omega = \pm v_m k^2 + \mathcal{O}(k^4)\,, \quad v_m = \sqrt{\frac{\lambda}{\chi_{ee}}\big(\chi_{ee}\chi_{nn}-\chi^2_{ne} \big)}\,.
\ee 
Finally, there is a pair of massive modes with a quadratic dispersion 
\be\begin{split}
\omega &= \pm m_0 \pm v_M k^2 +\mathcal{O}(k^4)\,, \\
v_M &= \frac{1}{2m_0} \Big( \lambda \chi_{\pi} + \big( \mu^2_0 \chi_{ee} - 2 \mu_0 \chi_{ne} +\chi_{nn} \big)\chi_v  \Big)\,.
\end{split}
\ee 
On the other hand, for the transverse sector we find 
\be \begin{split}
\omega_{\perp} =& \pm m_0 \sqrt{1+ \frac{\lambda_{\perp}}{\chi_v}k^2 } \\
\approx& \pm m_0 \pm \frac{\lambda_{\perp}}{2\chi_v} m_0 k^2\,.
\end{split}
\ee 

\section{Dissipative hydrodynamics}\label{sec:4}
In this section, we present a dissipative completion of the ideal theory and study (linearized) dissipative hydrodynamics of fracton superfluids. After postulating the derivative expansion, we systematically derive the most general hydrodynamic constitutive relations up to the third order in the derivative expansion by demanding consistency with the local form of the second law of thermodynamics. Then, we study the spectrum of the theory and verify the consistency of our approach. 

\subsection{Gradient expansion}

Before delving into the realms of dissipative hydrodynamics, we need to establish a systematic gradient counting scheme such that the hydrodynamic constitutive relations can be organized in a derivative expansion and the equations solved perturbatively. As we will shortly discuss, in the case of s-wave superfluids there are certain inevitable subtleties in the power counting that arise due to the lack of the common scaling symmetry. As a result, the derivative expansion introduced here is in a sense not a standard one and it is therefore advisable to verify that it indeed represents a consistent truncation of the hydrodynamic currents and equations of motion.

In order to address this problem, let us list out the (non-exhaustive) set of consistency conditions that we demand from a feasible derivative counting scheme. In hydrodynamics, the dispersion relations $\omega_i(k)$ are organized as a power series in the wavevector
\be \label{eq:expansion}
\omega_{i}(k) = \sum^{\infty}_{n=0} a_{ni} k^n_{i}\,.
\ee 
Here, the index $i$ represents the different hydrodynamic modes. Then, we require the following conditions:
\begin{enumerate}\label{eq:conditions}
    \item \textbf{Consistent expansion.} The $n$-th order coefficients $a_{ni}$ are uniquely determined by the $(n-1)$-th order constitutive relations. For example, the $0$-th order hydrodynamics exactly fixes the coefficients corresponding to terms linear in $k$. 
    \item \textbf{Stable modes.} All modes are linearly stable $\mathrm{Im} \,\omega_i \leq0$ on account of the constraints from the second law of thermodynamics and stability of the entropy density function\footnote{In general, the Landau instability may occur at the critical superflow destroying superfluidity. However, in this work we assume expansion around the state with no superflow.}. 
    \item \textbf{Dissipation-entropy correspondence.} The only transport coefficients that contribute to the attenuation of the modes are the ones that add to the entropy production. 
\end{enumerate}

We now introduce our derivative counting prescription. To this aim, let us recall the discussion below \eqref{eq:secondLawInternal} and \eqref{eq:josephsons} where we have established that the variables corresponding to gapped degrees of freedom ought to be counted as order one quantities. Therefore, the fluid variables are assigned the following orders\footnote{One could equivalently work with the invariant superfluid velocity $\xi_{ij} \sim \mathcal{O}(1)$ instead of the charge Goldstone or directly in terms of the two Goldstones $\theta \sim  \mathcal{O}(\nabla^{-2})$ and $\psi_i \sim \mathcal{O}(\nabla^{-1})$. See discussion below \eqref{eq:josephsons}.}
\be 
\theta \sim \mathcal{O}(\nabla^{-2})\,, \quad \{\psi_i\,, p_i\} \sim \mathcal{O}(\nabla^{-1})\,, \quad \{n\,, \epsilon\,, \Tilde{p}_i\,, \xi_{ij}\} \sim \mathcal{O}(1)\,, \quad \{\pi_i\,, v^s_i\} \sim \mathcal{O}(\nabla)
\ee 
while for the conjugate quantities we have 
\be \label{eq:conjugate}
\{\mu\,, T\,, \Tilde{V}_i\,, F_{ij}\} \sim \mathcal{O}(1)\,, \quad \{\mu_i\,, \lambda_i\} \sim \mathcal{O}(\nabla)\,.
\ee 
A full set of independent hydrodynamic equations capturing the dynamics of dissipative fracton superfluids is then given by
\be \begin{split}
 \label{eq:eoms}
\partial_t n + \partial_i J^i &= 0\,, \\
\partial_t \pi_i + \partial_j K^{ji} +J^i &= 0\,, \\
\partial_t \tilde p_i + \partial_j \tilde T^{ji} -f_i&= 0\,, \\
\partial_t \epsilon + \partial_i \epsilon^i &=0\,, \\
\partial_t  \psi_i  -\Big( \mu_i+\tilde \mu_i^d \Big)  &= 0  \,, \\
\partial_t \theta-\Big( \mu+\mu^d \Big) &=0\,.
\end{split}
\ee
Notice that we have decided to parameterize our fluid in terms of the independent variables $ \{n\,, \epsilon\,,  \tilde p_i\,, \theta\,, \psi_i\,, \pi_i\}$ while also allowing for the presence of arbitrary dissipative corrections to the Josephsons relations parameterized with $\tilde \mu_i^d$ and $\mu^d$.  In particular, we find it helpful to work with the thermal momentum \(\tilde{p}_i\), this variable accounts for the momentum carried by the (non-fractonic) thermal degrees of freedom. In terms of $\tilde p_i$ the momentum conservation reads
\be 
\partial_t \tilde p_i + \partial_j \tilde T^{ji} = f_i\,,
\ee 
where
\be
\tilde T^{ji} =T^{ji}+J^j u_i + K^{jk}\xi_{ik} 
\ee
is the invariant stress tensor, while 
\be 
f_i = n  \partial_t u_j + \pi_j\partial_t\xi_{ij} + J^k \partial_i v^s_k + K^{kj} \partial_k\xi_{ij}
\ee 
can be interpreted as a force. Once the constitutive relations are expressed in terms of the fluid variables the equations may be solved to a desired order in the derivative expansion.

Let us now ascertain the appropriate truncation scheme for the equations. Firstly, as has been pointed out below \eqref{eq:thermoIdentities}, the divergence of the dipole current appears in the equations akin to the charge current. Therefore, the dipole current needs to be expanded up to one order lower than the charge current, and the dipole equation is to be truncated at one order lower than that of charge conservation. Secondly, the dipole Goldstone $\psi_i$ is order minus one, hence the corresponding equation should be truncated at one order lower as compared with the equations for order zero variables. Finally, the charge Goldstone $\theta$ is actually order minus two and therefore the associated equation needs to be truncated at two orders lower. 
Therefore, the hydrodynamic equations are to be truncated as 
\be \begin{split}\label{eq:truncated}
\partial_t n + \partial_i J^i &= \mathcal{O}(\nabla^{n+2})\,, \\
\partial_t \pi_i + \partial_j K^{ji} +J^i &= \mathcal{O}(\nabla^{n+1})\,, \\
\partial_t \tilde p_i + \partial_j \tilde T^{ji} - f_i &= \mathcal O(\nabla^{n+2})\,, \\
\partial_t \epsilon + \partial_i \epsilon^i &=\mathcal{O}(\nabla^{n+2})\,, \\
\partial_t  \psi_i  -\Big( \mu_i+\tilde \mu_i^d \Big)  &= \mathcal{O}(\nabla^{n+1})  \,, \\
\partial_t \theta-\Big( \mu+\mu^d \Big) &=\mathcal{O}(\nabla^{n})\,.
\end{split}
\ee
From a practical standpoint, however, we find it convenient to parameterize our fluid in terms of the invariant superfluid velocity \(v^s_i\) rather than the dipole Goldstone \(\psi_i\). This choice is well-motivated on physical grounds, as this combination of Goldstone fields precisely corresponds to the massive degree of freedom. In terms of the new variables the set of invariant hydrodynamic equations read
\be \label{eq:eomsM}
\begin{split}
\partial_t n + \partial_i J^i &= \mathcal O(\nabla^{n+2})\,, \\
\partial_t \pi_i + \partial_j K^{ji} +J^i &= \mathcal O(\nabla^{n+1})\,, \\
\partial_t \tilde p_i + \partial_j \tilde T^{ji} - f_i &= \mathcal O(\nabla^{n+2})\,, \\
\partial_t \epsilon + \partial_i \epsilon^i &= \mathcal O(\nabla^{n+2})\,, \\
\partial_t  v^s_i  - \Big( \partial_i  \mu -  \mu_i +\mu^d_i \Big) &=  \mathcal O(\nabla^{n+1})  \,, \\
\partial_t \theta-\Big( \mu+\mu^d \Big) &=\mathcal O(\nabla^{n})\,,
\end{split}
\ee
where we have redefined the dissipative dipole chemical potential as $\mu_i^d=\partial_i \mu^d - \tilde\mu^d
_i$, and have truncated the equations accordingly. From now on, we will work with equations \eqref{eq:eomsM} as our set of hydrodynamic equations and $\{n\,, \epsilon\,, \tilde p_i\,, \pi_i\,, \theta\,, v^s_i \}$ as the hydrodynamic variables.

Gradient expansion in hydrodynamics is typically endowed with a power counting scheme, where a certain weight $z$ is assigned to time derivatives, i.e. $\partial_t \sim \mathcal O(\nabla^z)$. For example, in ordinary hydrodynamics one has $z=1$ whereas p-wave fracton superfluids exhibit $z=2$ \cite{PhysRevE.107.034142,Jain:2023nbf}. These scalings are consistent with the low-energy spectrum containing gapless modes with linear and quadratic dispersion, respectively. 

However, for s-wave fracton superfluids, the spectrum contains propagating modes with both soundlike $(\omega\sim k)$ and magnonlike $(\omega \sim k^2)$ dispersion relations (see Eq. \eqref{eq:sound} and \eqref{eq:mangon}). It is therefore not clear if one should count $\partial_t \sim \mathcal O(\nabla)$ or $\partial_t \sim \mathcal O(\nabla^2)$ as both of these counting schemes seem to be in conflict with the scaling of the magnonlike and soundlike mode, respectively. This issue was initially highlighted in \cite{Armas:2023ouk, Jain:2023nbf}. Intuitively, this tension can be attributed to the fact that the fractonic degrees of freedom vary in time more slowly that the neutral ones. Therefore, in the case of s-wave fracton superfluids it does not appear consistent to assign a common scaling to time derivatives\footnote{Unless certain coefficients are assigned anomalous scaling dimensions, though we do not explore this possibility here.}. Indeed, we have explicitly verified that assuming either $z=1$ or $z=2$ leads to the violation of at least one of the consistency conditions listed in \eqref{eq:conditions} and thus does not constitute a self-consistent power counting prescription.

In order to circumvent this issue we proceed to carry out the derivative expansion without assigning a definite scaling to the time derivatives. Our   approach to derivative expansion can be understood simply as a truncation of the hydrodynamic constitutive relations in spatial gradients according to Eq. \eqref{eq:eomsM}. Once truncated, the equations are to be solved for the time evolution of the hydrodynamic variables. Similarly, one can also truncate the entropy production equation Eq. \eqref{eq:linearEntropyProduction} in order to classify the dissipative contributions allowed by the second law of thermodynamics at a given order in the perturbative expansion.

 As we will see in the next subsections, this prescription produces dispersion relations that are organized systematically and satisfy all the consistency conditions (see Table \ref{tab:Table4}). Hence it constitutes a consistent derivative counting scheme for fracton superfluids. This is one of the key results of this paper.

\subsection{Entropy current analysis}
To identify an entropy current for the system, we start with the first law of thermodynamics
\be 
d\epsilon = \mu dn + \mu_i d \pi_i + T ds + \Tilde{V}_i d \Tilde p_i + \lambda_i d v^s_i + F_{ij} d \xi_{ij}\,.
\ee 
Using $\xi_{ij}=\partial_i \psi_j = \partial_i \partial_j \theta - \partial_i v^s_j$ and $\tilde p_i = p_i + n \partial_i \theta + \pi_j \partial_i \partial_j \theta - \pi_j \partial_i v^s_j$ we arrive at
\be \label{eq_firstlaw2}
d\epsilon = \bar \mu dn + \bar \mu_i d \pi_i + T ds + \Tilde{V}_i d  p_i + \lambda_i d v^s_i + n \tilde V_i d \partial_i \theta - \bar F_{ij} d \partial_i v^s_j + \bar F_{ij} d \partial_i \partial_j \theta \,.
\ee 
where we have introduced effective variables
\be 
\bar{\mu}= \mu + \Tilde{V}_i \partial_i \theta\,, \quad \bar{\mu}_i= \mu_i + \Tilde{V}_j \partial_j \partial_i \theta - \Tilde{V}_j \partial_j v^s_i \,, \quad  \bar{F}_{ij} = F_{ij} + \tilde{V}_i \pi_j \,.
\ee 
After a series of algebraic manipulations, we arrive at the following expression that determines the constitutive relations of the currents 
\be \begin{split}
\Delta &= \partial_i \Big( s^i - P \frac{\tilde V_i}{T}-\frac{1}{T} \epsilon^i + \frac{\bar \mu}{T} J^i + \frac{\bar \mu_j}{T} K^{ij}  + \frac{\tilde V_j}{T} T^{ij} +  \frac{\mu^d_j}{T} \bar F_{ij} - \frac{\bar F_{ij}}{T} \partial_j \mu^d + \partial_j \frac{\bar F_{ji}}{T} \mu^d - \frac{ \tilde V_i}{T} n \mu^d \Big) \\
&+\Big( \epsilon^i - \big(sT+\tilde V_j \tilde p_j\big) \tilde V_i + \mu \lambda_i + \mu_j F_{ij}\Big) \partial_i\frac{1}{T}   \\&+ \Big(J^i+\lambda_i\Big) \Big( \frac{ \mu_i}{T} - \partial_i \frac{ \mu}{T} - \frac{\tilde V_j}{T} \partial_j v^s_i \Big) 
- \Big(K^{ij}+F_{ij}\Big) \partial_i \frac{\bar \mu_j}{T}   \\&-\Big( T^{ij} - P\delta_{ij} - \tilde V_i \tilde p_j - F_{ik} \partial_j \partial_k \theta + F_{ik} \partial_j  v^s_k + J^i \partial_j \theta \Big)\partial_i \frac{\Tilde{V}_j}{T}  \\
&  -  \mu^d_i \Big( \frac{\lambda_i}{T} + \partial_j \frac{\bar F_{ij}}{T} \Big) +\mu^d \Big(\partial_i \frac{ n \tilde V_i}{T}      - \partial_i \partial_j \frac{\bar F_{ij}}{T}\Big)  \geq 0\,.
\end{split}
\ee 
For the ideal sector we impose $\Delta = 0$ finding 
\be \begin{split}\label{eq:0order}
J^i &= - \lambda_i\,, \\
K^{ij} &= - F_{ij}\,, \\ 
T^{ij} & = P \delta_{ij} + \Tilde{V}_i \Tilde{p}_j + \lambda_i \partial_j \theta + F_{ik} \xi_{jk} \,, \\
 E^i & =   \Tilde{V}_i (sT + \Tilde{p}_j \Tilde{V}_j ) - \mu \lambda_i - \mu_j F_{ij}\,,\\ 
S^i &= s \Tilde{V}_i \,.
\end{split}
\ee 
These constitutive relations are in full agreement with the Poisson bracket method \eqref{eq:ideal}. On the other hand, the dissipative currents are specified by 
\be \begin{split}
\Delta &=\mathcal E^i \partial_i\frac{1}{T} + \mathcal J^i \Big( \frac{ \mu_i}{T} - \partial_i \frac{ \mu}{T} - \frac{\tilde V_j}{T} \partial_j v^s_i \Big) -\mathcal T^{ij} \partial_i \frac{\Tilde{V}_j}{T} 
  \\& - \mathcal K^{ij} \partial_i \frac{\bar \mu_j}{T}  -  \mu^d_i \Big( \frac{\lambda_i}{T} + \partial_j \frac{\bar F_{ij}}{T} \Big) +\mu^d \Big(\partial_i \frac{ n \tilde V_i}{T}      - \partial_i \partial_j \frac{\bar F_{ij}}{T}\Big)  \geq 0\,.
\end{split}
\ee 
In the remainder of this work, we focus on small deviations from the stationary configuration \eqref{eq:entropyDensity} and investigate linearized hydrodynamics by expanding the constitutive relations up to the first order in perturbations. In this special case thermal momentum $\tilde p_i$ satisfies the conservation equation $\partial_t \tilde p_i + \partial_j \tilde T^{ji} = 0$ with $\tilde T^{ij} = T^{ij} - n_0 \partial_t \theta \delta_{ij}$.

Entropy production truncated accordingly, up to the second order in deviations, can be written as 
\be \begin{split}\label{eq:linearEntropyProduction}
\Delta =  \mathcal E^i \partial_i \frac{1}{T} + \mathcal J^i \Big(  \frac{\mu_i}{T} - \partial_i  \frac{\mu}{T} \Big) -\mathcal{\tilde{T}}^{ij} \partial_i \frac{\Tilde{V}_j}{T} 
   - \mathcal K^{ij} \partial_i  \frac{\mu_j}{T}  -  \mu^d_i \Big( \frac{\lambda_i}{T} + \partial_j  \frac{ F_{ij}}{T} \Big) -\mu^d    \partial_i \partial_j  \frac{F_{ij}}{T}  \geq 0\,.
\end{split} \ee
Before classifying the complete set of constitutive relations, let us notice that the constitutive relations for $\mathcal J^i$ and $\mu_d^i$ contain the following terms
\be \begin{split}
\mathcal J^i &= \beta \Big( \frac{\mu_i}{T} - \partial_i \frac{\mu}{T}\Big) + \dots \\
\mu_i^d &= \gamma \Big( \frac{\lambda_i}{T} + \partial_j  \frac{F_{ij}}{T} \Big)+ \dots
\end{split}
\ee 
where $\beta\,, \gamma \geq 0$ and the dots represents other contributions that are not relevant to the forthcoming argument. We thus see that the equations for the gapped degrees of freedom $v^s_i$ and $\pi_i$ take the form of the relaxation equations 
\be \begin{split}
\partial_t \pi_i + \beta \chi_{\pi} \pi_i + \dots &= 0 \,, \\
\partial_t v^s_i + \gamma \chi_v v_i^s + \dots &= 0 \,. \\
\end{split}
\ee 
Therefore, there are again two possibilities. The first possibility correspond to the pure hydrodynamic regime where the transport coefficients are order one quantities $\beta\,, \gamma \sim \mathcal{O}(1)$. In this regime, one can set $\partial_t \pi_i + \beta \chi_{\pi} \pi_i \approx \beta \chi_{\pi} \pi_i$ and $\partial_t v^s_i + \gamma \chi_v v_i^s \approx \gamma \chi_v v_i^s$ effectively integrating out the gapped degrees of freedom in analogy to the case of diffusion discussed in \ref{sec:pureHydro}.

When the transport coefficients are made perturbatively small, however, such that $\beta\,, \gamma \sim \partial_t$ the approximation is no longer valid as the two contributions are of the same order and the gapped degrees of freedom undergo independent dynamics. In this regime, the associated relaxation times are large and quantities $\pi_i$ and $v^s_i$ are ``almost" conserved. A comparable scenario, referred to as the \textit{spin dynamical regime}, has been explored in the context of spin hydrodynamics in \cite{Hongo:2021ona}.

While the latter possibility is of considerable interest, in the rest of this work, we will focus solely on the former, pure hydrodynamic regime, setting $\partial_t \pi_i=\partial_t v^i_s=0$ from now onwards. After integrating out the gapped degrees of freedom, we end up with four dynamical equations 
\be \label{eq:pureHydroEqs}
\begin{split}
    \partial_t n + \partial_i J^i &=\mathcal{O}(\nabla^{n+2}) \,, \\
    \partial_t \tilde p_i + \partial_j \tilde T^{ji} &=\mathcal{O}(\nabla^{n+2})\,, \\
    \partial_t \epsilon + \partial_i \epsilon^i &=\mathcal{O}(\nabla^{n+2}) \,, \\
    \partial_t \theta - \Big( \mu +\mu_d\Big)&=\mathcal{O}(\nabla^n)\,.  \\
\end{split}
\ee 
Supplemented with the two constraints, namely $J^i = -\partial_j K^{ji}$ and $\psi_i=\partial_i \theta$. The entropy production \eqref{eq:linearEntropyProduction} evaluated on-shell with respect to the constraints is then expressed as 
\be \begin{split}\label{eq:entropyProductionPureHydro}
\Delta = -\partial_j \Big( K^{ji} \frac{\mu_i}{T} - K^{ji} \partial_i \frac{\mu}{T} \Big) +  \mathcal E^i \partial_i \frac{ 1}{T} -\mathcal{\tilde{T}}^{ij} \partial_i \frac{\Tilde{V}_j}{T} 
   - \mathcal K^{ij} \partial_i  \partial_j \frac{\mu}{T}  -\mu^d    \partial_i \partial_j  \frac{F_{ij}}{T}  \geq 0\,.
\end{split} \ee
Therefore, after absorbing the total divergence term into the definition of the entropy current, we finally arrive at the desired form of the second law 
\be \begin{split}\label{eq:desired}
 \mathcal E^i \partial_i \frac{1}{T}  -\mathcal{\tilde{T}}^{ij} \partial_i \frac{\Tilde{V}_j}{T} 
   - \mathcal K^{ij} \partial_i  \partial_j \frac{\mu}{T}  -\mu^d    \partial_i \partial_j  \frac{F_{ij}}{T}  \geq 0\,.
\end{split} \ee
The dissipative currents will be constructed out of the 'non-hydrostatic' data whose classification is provided in the Table \ref{tab:Table3}.
\begin{table}[t]
    \centering
   \begin{tabular}{|l|c|c|c|}
   \hline
& First order & Second order & Third order  \\
 \hline
   Scalars & $\partial_i \frac{\tilde V_i}{T}$ &$\partial_i \partial_j \frac{F_{ij}}{T}\,, \partial^2 \mu\,, \partial^2 \frac{1}{T} $& $\partial^2 \partial_i \frac{\tilde V_i}{T}$\\
  Vectors   & $\partial_i \frac{1}{T}$ & $\partial_i \partial_j \frac{\tilde V_j}{T}$ &$\partial^2 \partial_i \frac{1}{T}$\\
      Tensors & $\partial_{\langle i} \frac{\tilde V_{j\rangle}}{T}$ &$\partial_{\langle i} \partial_{j \rangle} \frac{\mu}{T}\,, \partial_i \partial_j \frac{1}{T}$&$\partial^2 \partial_{\langle i} \frac{\tilde V_{j\rangle}}{T}\,, \partial_i \partial_j \partial_k \frac{\tilde V_k}{T}$\\
  \hline
 \end{tabular} 
    \caption{Non-hydrostatic data classification for the $s$-wave fracton superfluids in the regime of pure (linearized) hydrodynamics.}
    \label{tab:Table3}
\end{table}
We are now able to identify the constitutive relations order by order in the derivative expansion. 

For the first order hydrodynamics, corresponding to the truncation of the entropy production \eqref{eq:desired} at the second order in derivatives i.e. $\partial_\mu S^{\mu} = \mathcal{O}(\nabla) \mathcal{O}(\nabla)$ we find
\be 
\begin{split}\label{eq:1order}
\mathcal E_{(1)}^i &=  \alpha \partial_i \frac{1}{T}\,, \\
T_{(1)}^{ij} &= -\zeta \partial_k \frac{\tilde V_k}{T} \delta_{ij} - \eta \partial_{\langle i} \frac{\tilde V_{j \rangle}}{T}
\end{split}
\ee 
with $\alpha\,, \zeta\,, \eta \geq0$.

Moving to second-order hydrodynamics, we observe that truncating entropy production at the third order, such that $\Delta \sim \mathcal{O}(\nabla^2)\mathcal{O}(\nabla)$, does not impose inequality constraints on transport coefficients. This is because such terms can be combined into squares, as follows: $\Delta \sim \Big(\mathcal{O}(\nabla)+\mathcal{O}(\nabla^2)\Big)^2 + \mathcal{O}(\nabla^4)$. Therefore, after completing squares, the positivity of entropy production is ensured as long as the $\mathcal{O}(\nabla)\mathcal{O}(\nabla)$ coefficients are positive, and any mistake made during the completion of a square is absorbed to higher order. 

Nevertheless, there are still possible (nondissipative) corrections to the currents that one can add at this order in the derivative expansion, namely
\be 
\begin{split}\label{eq:2order}
\mathcal E_{(2)}^i &= c_1 \partial_i \partial_j  \frac{\tilde V_j}{T}\,, \\
T_{(2)}^{ij} &= c_1 \partial^2 \frac{1}{T} \delta_{ij} +c_2 \partial^2 \frac{\mu}{T} \delta_{ij} +c_3 \partial_j \partial_k \frac{F_{jk}}{T}\delta_{ij}\,, \\
\mathcal{K}_{(1)}^{ij} &= c_2 \partial_k \frac{\tilde V_k}{T} \delta_{ij}\,, \\
-\mu_{(1)}^d &= c_3 \partial_i \frac{\tilde V_i}{T}
\end{split}
\ee 
where we have already imposed Onsager reciprocity relations.

Finally, we consider third order hydrodynamics contributing to the entropy production truncated at the fourth order. At this order we are able to construct squares of the form $\mathcal{O}(\nabla^2)\mathcal{O}(\nabla^2)$ and therefore we expect to establish some inequality type constraints. Indeed, the most general constitutive relations for third order hydrodynamics are 
\be 
\begin{split}\label{eq:3order}
\mathcal E_{(3)}^i &=-\bar \alpha \partial^2 \partial_i \frac{1}{T}+a_1  \partial^2 \partial_i \frac{\mu}{T}+a_2 \partial_i \partial_j \partial_k \frac{F_{jk}}{T}\,, \\
T_{(3)}^{ij} &= \bar \zeta_1 \partial^2 \partial_k \frac{\tilde V_k}{T} \delta_{ij}+ \bar \zeta_2 \partial_i \partial_j \partial_k \frac{\tilde V_k}{T} + \bar \eta \partial^2 \partial_{\langle i} \frac{\tilde V_{j \rangle}}{T} \,, \\
\mathcal{K}_{(2)}^{ij} &= -\sigma_{||} \partial^2 \frac{\mu}{T} \delta_{ij} - \sigma_{\perp} \partial_{\langle i} \frac{\partial_{j \rangle} \mu}{T} +a_1 \partial^2 \frac{1}{T}\delta_{ij}+a_3\partial_k \partial_l \frac{F_{kl}}{T} \delta_{ij}\,, \\
-\mu_{(2)}^d &=  \xi \partial_i \partial_j \frac{F_{ij}}{T}+a_2 \partial^2 \frac{1}{T} + a_3 \partial^2 \frac{\mu}{T} \,.
\end{split}
\ee 
The second law of thermodynamics is then satisfied provided that the dissipative transport coefficients satisfy $\bar \alpha\,, \bar \zeta_1\,, \bar \zeta_2\,, \bar\eta\,, \sigma_{||}\,, \sigma_{\perp}\,, \xi \geq 0$ and $\bar \alpha \sigma_{||} \geq  a^2_1$.
\subsection{Dispersion relations}
In this section we compute the dispersion relations $\omega(k)$ of the hydrodynamic modes up to the $\sim k^4$ contributions associated with the third order hydrodynamics. We confirm that the dispersion relations are organized systematically in a derivative expansion and that the constraints from the second law of thermodynamics and thermodynamic stability of the equilibrium state guarantee the linear stability of the modes.

Plugging the linearized constitutitve relations Eqs. \eqref{eq:0order}, \eqref{eq:1order}, \eqref{eq:2order} and \eqref{eq:3order} into the equations of motion \eqref{eq:pureHydroEqs} we arrive at the following set of dynamical equations for the third order hydrodynamics
\be \begin{split}
 \label{eq:Linearizedeoms}
\partial_t n + \partial_i \partial_j F_{ij} -\frac{c_2}{T_0}  \partial^2 \partial_i \tilde V_i +\sigma \partial^4 \frac{\mu}{T}-a_1 \partial^4 \frac{1}{T} -\frac{a_3}{T_0} \partial^2 \partial_i \partial_j F_{ij}&= 0\,, \\
\partial_t  \Tilde p_i + s_0 \partial_i  T - \frac{\hat \eta}{2T_0} \partial^2 \Tilde{V}_i - \frac{1}{T_0} \Big( \hat \zeta   + \hat \eta \frac{d-2}{2d}  \Big) \partial_i \partial_j \Tilde{V}_j +c_1 \partial_i \partial^2 \frac{1}{T} + c_2 \partial_i \partial^2 \frac{\mu}{T} + \frac{c_3}{T_0} \partial_i \partial_j \partial_k F_{jk} &=0 \\
\partial_t \epsilon + s_0 T_0 \partial_i \tilde V_i + \mu_0 \partial_i \partial_j F_{ij}  + \hat \alpha \partial^2 \frac{1}{T} + \frac{c_1}{T_0} \partial^2 \partial_i \tilde V_i+ a_1 \partial^4 \frac{\mu}{T} + \frac{a_2}{T_0} \partial^2 \partial_i \partial_j  F_{ij}  &= 0\,, \\
\partial_t \theta- \mu + \frac{c_3}{T_0} \partial_i \tilde V_i +\frac{\xi}{T_0}\partial_i \partial_j F_{ij} +\frac{a_2}{T_0}\partial^2 \frac{1}{T} + a_3 \partial^2 \frac{\mu}{T}&=0\,.
\end{split}
\ee
where we have defined the hatted coefficients representing corrected first order coefficients as follows
\be \begin{split}
\hat \eta &= \eta - \bar \eta \partial^2 \,, \\
\hat \zeta &= \zeta -  \big(\bar\zeta_1 +\bar\zeta_2\big)\partial^2 \,, \\
\hat \alpha &= \alpha-\bar\alpha \partial^2\,.
\end{split}
\ee 
Using the form of the entropy density \eqref{eq:entropyDensity} evaluated on the constraint $\psi_i=\partial_i\theta$ such that $\xi_{ij} = \partial_i \partial_j \theta$ we obtain the following thermodynamic identities  
\be \begin{split}
F_{ij} &= \lambda^{ijkl} \partial_k \partial_l \theta \,, \\
\delta \frac{1}{T} &=\frac{1}{T_0} \Big(   \chi_{ne} \delta n-\chi_{ee} \delta \epsilon\Big)\,, \\
\delta \frac{\mu}{T} &=\frac{1}{T_0} \Big(   \chi_{nn} \delta n-\chi_{ne} \delta \epsilon\Big)\,, \\
\tilde V_i &= \chi_p \tilde p_i\,, \\
 \delta T &=  T_0 \chi_{ee} \delta e - T_0 \chi_{ne} \delta n \,, \\
 \delta \mu &=  \chi_e \delta e +  \chi_n \delta n\,.
\end{split}
\ee
After expressing the variations of thermodynamic variables in terms of hydrodynamic densities using the above identities, we arrive at a closed system of differential equations for the evolution of the hydrodynamic variables
\be \begin{split}
 \label{eq:eomsFinal}
\partial_t \delta n + \lambda\partial^4 \theta -c_2\frac{\chi_p}{T_0}\partial^2 \partial_i \tilde p_i +  \tilde  \sigma \partial^4 \delta n+\tilde a_1 \partial^4 \delta \epsilon - \lambda \frac{a_3}{T_0} \partial^6 \theta &= 0\,, \\
\partial_t \tilde p_i +\hat\chi_{ee}\partial_i \delta \epsilon -\hat\chi_{ne} \partial_i \delta n - \frac{\chi_p }{2 T_0} \hat \eta  \partial^2 \Tilde{p}_i - \frac{\chi_p}{T_0} \Big( \hat \zeta   + \hat \eta \frac{d-2}{2d}  \Big) \partial_i \partial_j \Tilde{p}_j  + c_3 \frac{\lambda}{T_0} \partial^4 \partial_i\theta&= 0\,, \\
\partial_t \delta \epsilon +  \tilde \chi_p 
\partial_i \tilde p_i +\hat \lambda \partial^4 \theta + \hat \alpha_n  \partial^2 \delta n-\hat \alpha_e \partial^2 \delta \epsilon + c_1 \frac{\chi_p}{T_0} \partial^2 \partial_i \tilde p_i  &= 0\,, \\
\partial_t \theta- \hat \chi_n \delta n - \hat \chi_e \delta \epsilon +c_3 \frac{\chi_p}{T_0} \partial_i \tilde p_i +\frac{\lambda}{T_0} \xi \partial^4 \theta&=0\,.
\end{split}
\ee
For brevity of presentation, we have introduced the following notation
\be \begin{split}
\tilde \sigma &= \frac{1}{T_0}\Big(\sigma \chi_{nn}-a_1 \chi_{ne} \Big) \,, \\
\tilde a_1 &= \frac{1}{T_0}\Big(a_1 \chi_{ee}-\sigma \chi_{ne} \Big) \,, \\
\hat \lambda &= \lambda \Big( \mu_0 +\frac{a_2}{T_0} \partial^2 \Big)\,, \\
\tilde a_3 &= \frac{1}{T_0} \Big( a_3 \chi_{nn}-a_2 \chi_{ne} \Big) \,, \\
\tilde a_2 &= \frac{1}{T_0} \Big( a_2 \chi_{ee}-a_3 \chi_{ne} \Big) \,, \\
\hat \chi_{ee} &= s_0 T_0 \chi_{ee}-\frac{1}{T_0}\Big(c_1 \chi_{ee}+c_2 \chi_{ne}\Big)\partial^2 \,, \\
\hat \chi_{ne} &= s_0 T_0 \chi_{ne} -\frac{1}{T_0} \Big(c_1 \chi_{ne} +c_2 \chi_{nn}\Big)\partial^2\,, \\
\hat \alpha_n &= \frac{1}{T_0}\Big(  \chi_{nn} a_1\partial^2 + \chi_{ne} \hat \alpha \Big)\,, \\
\hat \alpha_e &= \frac{1}{T_0}\Big(\chi_{ne} a_1 \partial^2 + \chi_{ee} \hat \alpha   \Big)\,, \\
\hat \chi_n &=\chi_n-\tilde a_3\partial^2\,, \\
\hat \chi_e &= \chi_e-\tilde a_2\partial^2\,.
\end{split}
\ee 
Solving the eigenvalue problem, we verify the existence of two ``sound" modes in the longitudinal sector and one shear mode in the transverse sector with the exact formulas for the dispersion relations given by
 \be\begin{split}\label{eq:modesFull}
\omega_{\text{sound}} &= \pm\left( v_s  - \tilde v_s k^2\right)k- i\left( \Gamma + \tilde \Gamma k^2\right)k^2+\mathcal{O}(k^5) \,, \\
\omega_{\text{fracton}} &= \pm\left( v_m k + \tilde v_m k^3\right)k -i\Omega k^4 + \mathcal{O}(k^5)\,, \\
\omega_{\text{shear}} &= -i \left[\eta \frac{\chi_p}{2T_0} +\bar \eta \frac{\chi_p}{2T_0} k^2\right]k^2 +\mathcal{O}(k^5)\,.
\end{split}
\ee  
We have used tilded constants to denote the subleading corrections to the velocities and attenuation constants of the modes. 
\begin{table}[t]
    \centering
   \begin{tabular}{|l|c|c|c|c|}
   \hline
 & Zeroth order & First order & Second order & Third order  \\
 \hline
   $\omega_{\text{sound}}$ & $\pm v_s k$ & $-i\Gamma k^2$ & $\mp\tilde v_s k^3$& $-i\tilde \Gamma k^4$ \\
  $\omega_{\text{fracton}}$  & - & $\pm v_m k^2$ & - & $\pm \tilde v_m k^4-i\Omega k^4$ \\
      $\omega_{\text{shear}}$ & - & $-i\eta \frac{\chi_p}{2T_0}k^2$ & - & $-i\bar \eta \frac{\chi_p}{2T_0}k^4$\\
  \hline
 \end{tabular} 
    \caption{Contribution of momentum as polynomial expressions to the hydrodynamic modes in $s$-wave fracton superfluids, determined at a specific order of the derivative expansion.}
    \label{tab:Table4}
\end{table}
The explicit expressions for the constants are listed below
\be \begin{split}
v_s&=s_0 T_0 \sqrt{ \chi_{ee}  \chi_p }\,, \\
 v_m &= \sqrt{\frac{\lambda}{\chi_{ee}}\Big(\chi_{ee}\chi_{nn}-\chi^2_{ne} \Big)}\,, \\
\Gamma &= \frac{1}{2T_0 } \Big(  \big( \zeta + \eta \frac{d-1}{2d}\big) \chi_p + \alpha \chi_{ee} \Big) \,, \\
\tilde v_s &= \frac{1}{2 v_s} \Big[\Big(\Gamma - \frac{\alpha \chi_{ee}}{T_0}\Big)^2 -\frac{\lambda}{\chi_{ee}}\Big(\mu_0 \chi_{ee}-\chi_{ne} \Big)^2 \Big]\,, \\
\tilde \Gamma &= \bar \alpha \frac{\chi_{ee}}{2 T_0}+\Big(\sigma_{||}+\sigma_{\perp}\frac{d-1}{d} \Big) \frac{  \chi _{ne}^2}{2 T_0 \chi _{ee}}+\Big(\bar \zeta_1 +\bar \zeta_2+\bar \eta \frac{d-1}{d} \Big)\frac{\chi_p}{2T_0} -a_1 \frac{ \chi_{ne}}{T_0} \\
&-\Big(\zeta+\eta \frac{d-1}{d} \Big) \frac{  \chi_p \lambda  \left(\chi_{ne}-\mu _0 \chi_{ee}\right)^2}{2T_0 v^2_s \chi_{ee} }\,, \\ 
\Omega &= \xi \frac{\lambda}{2T_0}+\Big(\sigma_{||}+\sigma_{\perp}\frac{d-1}{d} \Big)\frac{\chi_{nn}\chi_{ee}-\chi^2_{ne}}{2T_0\chi_{ee}}+\Big(\zeta+\eta \frac{d-1}{d} \Big) \frac{  \chi_p \lambda  \left(\chi_{ne}-\mu _0 \chi_{ee}\right)^2}{2T_0 v^2_s \chi_{ee} }\,, \\
\tilde v_m &=  v_m \Big(\frac{a_3}{T_0} - \frac{  \lambda  \left(\chi_{ne}-\mu _0 \chi_{ee}\right)^2}{2 v^2_s \chi_{ee} } \Big)\,.
\end{split}
\ee 
It is straightforward to verify that the dispersion relations satisfy the consistency conditions \eqref{eq:conditions}. From the Table \ref{tab:Table4} it follows that the dispersion relations are organized consistently with respect to the derivative counting scheme in the sense that the transport coefficients corresponding to the $n$-th order hydrodynamics contribute to the $k^{n+1}$ order in the power series expansion of the dispersion relations $\omega(k)$ (consistent expansion). Furthermore, the conditions for thermodynamic stability Eq. \eqref{eq:stability} and second law of thermodynamics\footnote{These are the constraints on transport coefficients listed below Eqs. \eqref{eq:1order} and \eqref{eq:3order}.} guarantee that the imaginary part of the modes is negative so that the modes are linearly stable throughout the free parameter space (stable modes). We can also see that the only transport coefficients that contribute to the attenuation of the modes are the ones that produce entropy. In particular, the nondissipative coefficients $c_1, c_2$ and $c_3$ do not contribute to the attenuation of the modes (dissipation-entropy correspondence).

 \section{Discussion}\label{sec:discussion}
Motivated by the theoretical and experimental pursuit of fracton phases of matter, we have developed a hydrodynamic theory for dipole-conserving many-body systems with spontaneously broken U(1) symmetry. In this section, we summarize our results, put them in a broader perspective, and propose some promising future directions. 

We started our study by constructing the hydrodynamic theory for simple systems with intrinsic dipole moments. This was done by incorporating a non-hydrodynamic degree of freedom corresponding to the internal dipole density $\pi_i$. Our construction is, in that regard, analogous to the theories of spin hydrodynamics \cite{Gallegos:2021bzp,Hongo:2021ona,Gallegos:2022jow}, wherein the role of intrinsic dipole is played by the spin density $\sigma_{ij}$. We noted an existence of two distinct counting prescriptions corresponding to different physical regimes: the dipole dynamical regime, where the internal dipole density is an independent dynamical quantity, and the pure hydrodynamic regime, where it can be considered as effectively relaxed onto a local equilibrium state.

Then we have analyzed the zero-temperature theory of fracton superfluids. In particular, starting from the $p$-wave phase we have shown how to break the U(1) symmetry arriving at the effective Goldstone description for the $s$-wave superfluid phase. The resulting theory has an interpretation of a two-component fluid consisting of the charge condensate coupled to the dipole condensate. We provide a hydrodynamic interpretation of the model and confirm the existence of two modes: a gapless scalar mode $\omega \sim k^2$ and a massive vector mode $\omega \sim m + k^2$. 

Next we have generalized the Goldstone theory to a finite-temperature regime. Importantly, the finite temperature theory contains a neutral component specified by the invariant momentum $\tilde p_i$, which is not charged under monopole and dipole symmetry. The origin of the neutral component is not fractonic in nature in the sense that it behaves just like the conventional uncharged matter. The hydrodynamic analysis of this section was done using the method of Poisson bracket and hence limited to the ideal sector. The spectrum of the theory contained an additional sound mode with the usual linear dispersion $\omega \sim k$.  

Finally, we have applied the entropy current formalism and performed an analysis of dissipative superfluids. This required establishing a consistent derivative counting prescription, which proved to be a non-trivial task given the incompatible nature of the normal and fractonic components. Indeed, after trying the two different approaches proposed in \cite{Armas:2023ouk,Jain:2023nbf} we have eventually arrived at the conclusion that neither counting is systematic due to the lack of the universal scaling symmetry. Subsequently, we have carried out a derivative expansion in a way that is agnostic of the scaling of the time derivatives and confirmed the consistency of the expansion by computing the dispersion relations for third order hydrodynamics.

We now point out some interesting directions for future research. First, it would be interesting to study in detail the hydrodynamic two point functions to further confirm the self-consistency of the derivative expansion scheme. In addition, in the studies of dissipative
superfluids we have integrated out all the gapped degrees of freedom and focused solely on the pure hydrodynamic regime. It would be interesting to explore various quasihydrodynamic regimes where either the internal dipole, the massive Goldstone\footnote{This scenario has been partially studied in \cite{Armas:2023ouk}, where a small parameter was attached to the mass of the Goldstone.}, or both are dynamical. This can be done by rendering certain transport coefficients, related to the gaps of the modes, perturbatively small as we have done in the dipole dynamical regime. 

Given the large number of tensor structures allowed in the most general setting, we have restricted dissipative considerations to the study of small fluctuations around the stationary configuration with vanishing neutral momentum $\tilde{p}^0_i = 0$ and zero superflow $\xi^0_{ij} = 0$. A natural extension of this work thus include a derivation of the nonlinear dissipative constitutive relations. A large value of the equilibrium superflow generally leads to a Landau instability destroying superfluidity (see \cite{PhysRevD.108.L081903,Arean:2023nnn} for modern treatment). Therefore, what appears especially interesting to the authors is the study of the critical superflows in fractonic superfluids. To this aim one needs to derive the hydrodynamic constitutive relations linearized around a finite superflow $\xi_{ij}=\xi^0_{ij}$ and study the stability of the modes. 

Finally, our construction paves the way towards a hydrodynamic theory of crystals with topological defects (hydrodynamics of crystals with non-topological defects have been worked out in \cite{mabillard_nonequilibrium_2021}) in terms of fractonic degrees of freedom via application of fracton-elasticity duality. Topological defects have been identified as the source of plasticity, near thermal equilibrium (see e.g. \cite{PhysRevB.107.155108}). Moreover, they have been associated with the weak breaking of emergent higher-form symmetries \cite{PhysRevE.105.024602,Armas:2023tyx}. It would be interesting to understand these properties from the fracton side of the duality. Dislocations, constituting a class of topological defects, correspond to dipoles in the dual theory. They are typically constrained to move along their respective Burger vectors. This is known as the glide constraint. The way to implement that constraint within hydrodynamic construction would be to additionally impose the quadrupole symmetry, which would fix the motion of the dipoles to one spatial dimension. Therefore, the next step in generalizing our construction, motivated by the fracton-elasticity duality, would be to implement both dipole and quadrupole symmetry into the hydrodynamic description. 
\section*{Acknowledgements}\label{sec:ack}
We thank the anonymous referee for their valuable comments.

A.G. and P.S. have been supported, in part, by
the Polish National Science Centre (NCN) Sonata Bis
Grant 2019/34/E/ST3/00405. F.P.-B. has received funding
from the Norwegian Financial Mechanism 2014-2021 via the
NCN, POLS Grant 2020/37/K/ST3/03390. A.G. was supported through a stipend from the International Max Planck Research School (IMPRS) for Quantum Dynamics and Control hosted at the Max Planck Institute for the Physics of Complex Systems.

 \bibliographystyle{JHEP}
\bibliography{biblio.bib}

\end{document}